\documentstyle[aps,prl,multicol,epsfig]{revtex} 
\begin{document}
\title{ Quantum fluctuation generated vortices, dual singular gauge transformation and
       zero temperature transition from d-wave superconductor to 
       underdoped regime }
\author{ Jinwu Ye }
\address{ Department of Physics and Material Research Institute, The Pennsylvania State University, 16802 }
\date{\today}
\maketitle
\begin{abstract}
  By extending the original Anderson singular gauge transformation for static vortices
  to two mutual flux-attaching singular gauge transformations for moving vortices, 
  we derive an effective action describing the zero temperature quantum phase transition
  from d-wave superconductor to underdoped regime. In this action, quantum fluctuation
  generated vortices couple to quasi-particles by a mutual statistical interaction with
  statistical angle $ \theta= 1/2 $ and a dynamic Doppler shift term, the vortices are also
  interacting with each other by long-range logarithmic interactions due to charge fluctuation.
  Neglecting the charge fluctuation first, we find that the mutual statistical interaction is
  exactly marginal. In the underdoped regime, the quasi-particles are described by $2+1$ dimensional
  QED; in the superconducting regime, they are essentially free. However, putting back the charge
  fluctuation changes the physical picture dramatically: both the dynamic Doppler shift term and the
  mutual statistical interaction become {\em irrelevant} short-ranged interactions on both sides of
  the quantum critical point. There are no spin-charge separation and {\em no} dynamic gapless gauge
  field in the Cooper-pair picture. The formalism developed at $ T=0 $ is applied to study thermally
  generated vortices in the vortex plasma regime near the finite temperature KT transition. The important
  effects of the AB phase scattering and the Doppler shift on ARPES data presented in Phys. Rev. Lett.
  87, 227003 (2001) are also briefly reviewed. 

\end{abstract}
\begin{multicols}{2}
\section{ Introduction}

   In this paper, we are trying to study the nature of zero 
  temperature quantum phase transition from d-wave superconductor
  at $ x > x_{c} $ to the underdoped regime at $ x < x_{c} $ of
  the high temperature superconductors ( Fig.1).

\vspace{0.25cm}

\epsfig{file=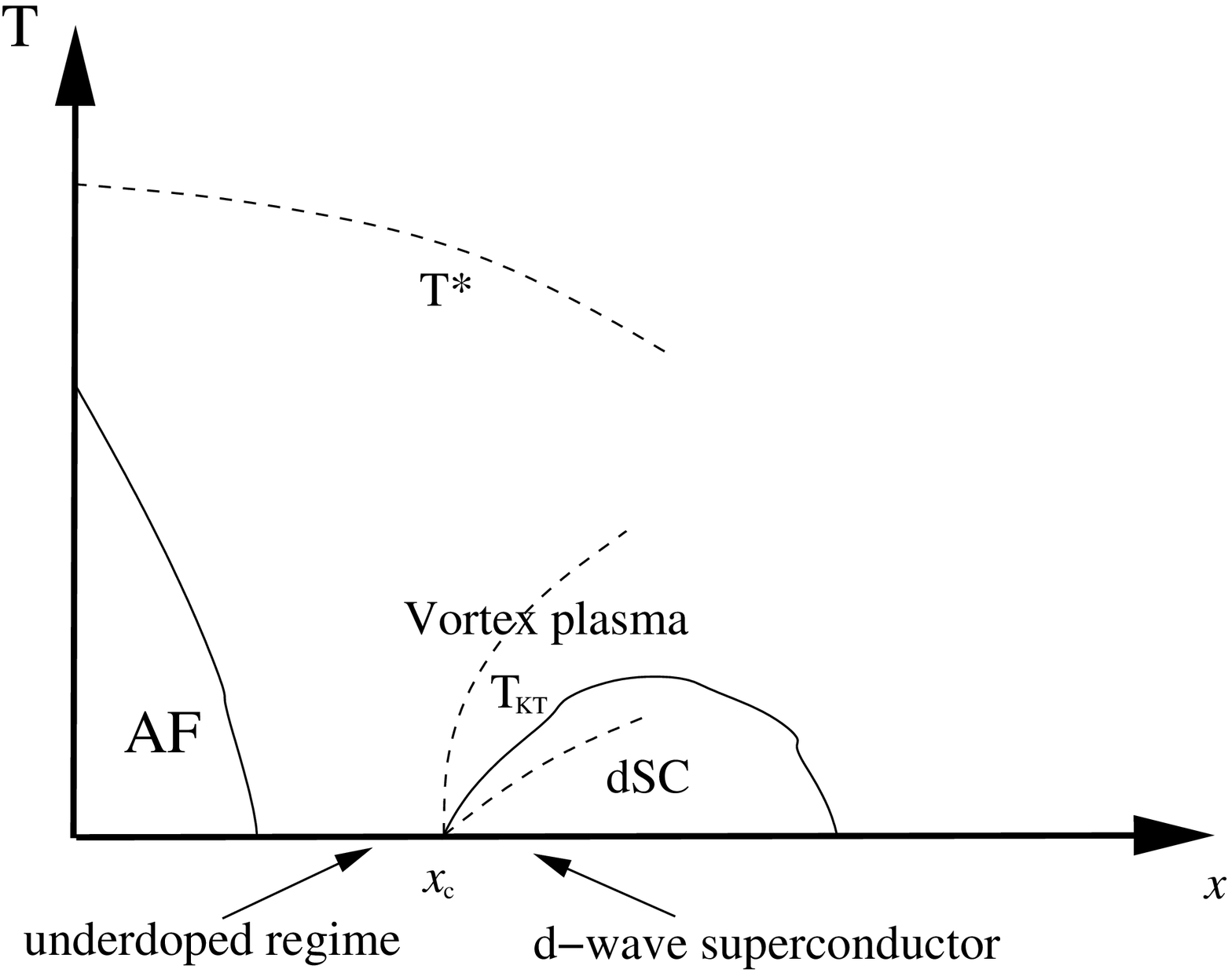,width=3.2in,height=2in,angle=0}

{\footnotesize {\bf Fig 1:} The temperature ( T ) versus doping ( $ x $ )
  diagram of high $ T_{c} $ cuprates.}

\vspace{0.25cm}

    It is well known that inside the superconductor phase $ x>x_{c} $,
  there are low energy
  quasi-particle excitations near the four nodes of the Fermi surface,
  the quantum phase fluctuation of the Cooper pair condensates are suppressed,
  the positive and negative vortices are bound together.
  As the doping decreases from the right to $ x_{c} $, the phase fluctuation
  increases. At the underdoped regime
  $ x < x_{c} $, the quantum phase fluctuations are so strong that
  they generate free $ hc/2e $ vortices, destroy the long range
  phase coherence of the d-wave superconductor, 
  therefore the superconducting ground state.
  However, the local short-range pairing still exists, the low energy
  quasi-particles at the four nodes remain.
  This quantum phase transition is driven by the condensation of
  $ hc/2e $ vortices. Starting from this physical picture, Balents, Fisher and Nayak discussed Quantum phase
  transitions from $d$-wave superconductor
  to underdoped regime and to the anti-ferromagnet regime with focus on the condensation of $ hc/e $ vortices
  in \cite{balents}. Earlier discussions on the stability of $ hc/e $ vortex starting
  from Mott insulating phase were given in Ref.\cite{own}.
  The present paper focus on the condensation of $ hc/2e $ vortices. 

  Mathematically, we can describe the transition by starting from the
  $ d $ wave BCS Hamiltonian inside the superconducting state $ x > x_{c} $
  in the presence of external magnetic field $ \vec{A} $ \cite{lee}:
\begin{equation}
      H=\int d x
      d^{\dagger}(x) 
     \left ( \begin{array}{cc}
		h + V(x)  &   \hat{\Delta}   \\
		\hat{\Delta}^{\dagger}   &   -h^{*} -V(x)  \\
		\end{array}   \right )  d(x)
\label{start}
\end{equation}
   with $ d_{\uparrow}(x)= c_{\uparrow}(x), d_{\downarrow}(x)
     = c^{\dagger}_{\downarrow}(x) $ \cite{exchange},
   $ V(x) $ is random chemical potential due to {\em non-magnetic} impurities
   which will be set to be zero in the following.

   Based on the earlier work on $d$-wave superconductor by Simon and Lee \cite{lee} and on
   $ t- J $ model by Ye and Sachdev \cite{deri}, 
   the author derived the gauge-invariant form in Ref.\cite{static}:
\begin{eqnarray}
  h   & = & \frac{1}{2m} (\vec{p}-\frac{e}{c} \vec{A} )^{2} -\epsilon_{F}   \nonumber    \\
  \hat{\Delta} &=& \frac{1}{4 p_{F}^{2}}[
  \{p_{x}-\frac{1}{2} \partial_{x}\phi, p_{y}-\frac{1}{2} \partial_{y} \phi\}
     \Delta(\vec{r})
             \nonumber  \\
  &+& (p_{x}-\frac{1}{2} \partial_{x} \phi) \Delta(\vec{r}) 
  (p_{y}+\frac{1}{2} \partial_{y} \phi)
         \nonumber  \\
  &+&(p_{y}-\frac{1}{2} \partial_{y} \phi) \Delta(\vec{r}) 
   (p_{x}+\frac{1}{2}\partial_{x} \phi)
                    \nonumber  \\
   & + & \Delta(\vec{r}) \{p_{x}+\frac{1}{2} \partial_{x} \phi, 
   p_{y}+\frac{1}{2}\partial_{y} \phi \} ] 
\end{eqnarray}
  where $\phi $ is the phase of the order parameter $ \Delta(\vec{r}) $ and
  $ p_{x} $ and $ p_{y} $ are the momenta relative to the origin.

   Because high $ T_{c} $ cuprates are strongly type II superconductors
  ( $ \kappa=\lambda/\xi \gg 1 $ ), we assume
  the gap amplitude to be a constant $ \Delta(x) \sim \Delta_{0}
   e^{i \phi} $ across the transition from $ x > x_{c} $ to $ x < x_{c} $.
  The phase $ \phi $ can be set to be zero inside the superconducting
  state $ x > x_{c} $
  because its fluctuation is suppressed. However in the underdoped
  regime $ x < x_{c} $, there are strong phase fluctuation which leads to
  the free $ hc/2e $ vortices located at
  $ \vec{R}_{i} $ with parity $ n_{i} = \pm 1 $, then the phase $ \phi $
  satisfies $ \nabla \times \nabla \phi=2 \pi \hat{z} \sum_{i} n_{i}
  \delta(\vec{r}-\vec{R}_{i} ) $.

  A lot of work has been done on a closely related problem where
 quasi-particles are coupled to the vortices generated by external
 magnetic field inside the superconducting state 
 \cite{lee,static,mac,volovik,hirs,and,sing,bert1,india,kov1,kov2}. 
 Employing semi-classical approximation, Volovik pointed out that  
 the circulating supercurrents around vortices
 induce Doppler energy shift to the quasi-particle spectrum, which leads to
 a finite density of states at the nodes \cite{volovik}. 
 This effect ( Volovik effect) has been employed to
 attempt to explain the experimental observations
  \cite{ong} of longitudinal thermal
 conductivity $ \kappa_{xx} $ in \cite{hirs}.

 Starting from BCS Hamiltonian, Wang and MacDonald performed
 a first numerical calculation on quasi-particle spectrum in vortex lattice
 state \cite{mac}.  By phenomenological scaling arguments,
 Simon and Lee (SL) proposed
 the approximate scaling forms for longitudinal and transverse thermal conductivities
 for {\em dirty} $ d $ wave superconductors in the mixed state \cite{lee}.
 Anderson \cite{and} employed the
 first single-valued singular
 gauge transformation to study quasi-particle dynamics in the mixed state.
 Unfortunately, Anderson
 made an incorrect mean field approximation which violates the "Time-Reversal"
 symmetry \cite{static}, therefore leads to the incorrect conclusion that
 there is Landau level quantization of energy levels of the quasi-particle. 
 Franz and Tesanovic employed a different  single-valued 
 singular gauge transformation and studied the quasi-particle spectrum numerically \cite{sing}.
 They did not see the signature of Landau level quantization in their numerical
 calculations. More detailed studies in various kinds of vortex lattice states
 were conducted in Refs.\cite{bert1,india}. The quasi-particle spectrum around a single vortex, paying
 special attention to the strong anisotropy of Fermi and gap velocities, was given in \cite{kov1,kov2}. 

 In the vortex lattice state, the vector potential only
 provides a periodic potential instead of scattering quasi-particles,
 its effect on the energy spectrum and other physical quantities
 is not transparent. In Ref.\cite{static}, the author studied the quasi-particle transport
 in the disordered vortex state where the random AB
  phase scattering may show its important effects.
  I observed that because infinite thin $ hc/2e  $
  vortices do {\em not} break the T reversal symmetry \cite{exchange},
  there is no Landau level quantization, any correct mean field theory should respect this T symmetry (The other way around
  need not be necessarily true, for example, in $ \nu=1/2 $ Fractional Quantum Hall system\cite{hlr}, the mean field state is a Fermi
  surface of composite fermion which respects T symmetry even the original electron system breaks the T symmetry due to
  the external magnetic field).
  I applied the Anderson singular transformation Eqn.\ref{first} to {\em disordered}
  vortex state and found that
  the long-range logarithmic interaction between vortices suppresses the
  fluctuation of superfluid velocity (scalar potential), but does {\em not}
   affect the fluctuation of the internal gauge field.
   Therefore the scalar field acquires a " mass " determined by the vortex
   density, but the gauge field remains  " massless ".
  The quasi-particle scattering from the "massless" internal
  gauge field dominates over those from the well-known "massive"
  Volovik effect and the non-magnetic scattering
  at sufficient high magnetic field. This dominant scattering mechanism
   is a purely quantum mechanical effects which
   was overlooked by all the previous semi-classical treatments \cite{hirs}.  In fact. it is responsible for the
  behaviors of both $ \kappa_{xx} $ and $ \kappa_{xy} $ in high magnetic field
  observed in the experiments \cite{ong}.  

     When the vortices are generated by quantum fluctuations 
 themselves are moving around, new physics may
 arise. For example, moving $ hc/2e $ vortices does break T reversal
  symmetry and leads to Hall voltage drop, a moving $ hc/2e $ vortex is different from a moving
  $ - hc/2e $ vortex. This in-equivalence renders FT gauge in Ref.\cite{sing} useless for moving vortices although
  it is equivalent to Anderson's gauge for static vortices. Anderson gauge has to be employed to study moving vortices.
 Intuitively, from Newton's third law, the vortices must also feel the counter-acting AB
 phase coming from the quasi-particles. Therefore there is a mutual AB phase
 scattering between vortices and quasi-particles. The dynamic Doppler shift term on the quasi-particles
 can also be equivalently viewed as an effective gauge field coupled to the phase fluctuation (see Eqn.\ref{act}).
  In this paper, we enlarge Anderson singular gauge transformation
 to {\em two} mutual flux-attaching singular gauge transformations. We then apply them to study the physics of
 quasi-particles coupled to moving vortices near the zero temperature quantum critical point.
  In canonical quantization  and in Coulomb gauge,
  we perform the two singular transformations (which are dual to each other)
  to quasi-particles(which are spinons) and moving vortices(which are holons) respectively to satisfy all
   the possible commutation relations. 
  Just like a conventional singular gauge transformation leads to a
  conventional Chern-Simon (CS) term, the two mutual singular gauge
  transformations lead to a mutual CS term. This elegant mutual CS term 
  describe the mutual AB phase scattering between vortices and quasi-particles.
  Alternatively, in path-integral presentation, the effective action describes
 the quasi-particle moving
 in both vector and scalar potentials due to the phase fluctuations of
 quantum generated vortices. By a duality transformation presented
  in Refs. \cite{dual,berry} to a vortex representation, the 
  quantum fluctuation generated vortices couple to quasi-particles
  by a mutual CS term, the vortices are also interacting with
  each other by long range logarithmic interactions due to charge fluctuation 
  described by a Maxwell term. 
  Neglecting the charge fluctuation and its associated dynamic Doppler shift term first, we find a fixed line
  characterized by the mutual statistical angle $ \theta $ and
  calculate the universal
  spinon, vortex and mutual Hall drag conductivities
  which continuously depend on $ \theta $.
 This transition can also be viewed as a simple example of confinement and deconfinement transition.
  In the disordered phase, the quasi-particles are described by $2+1 $ dimensional QED.
  In the superconducting side, they are essentially free. We stressed explicitly the lack of
  periodicity of $ U(1) $ mutual CS gauge theory.
  When the $ U(1) $ charge fluctuation is taken into account,
  we treat both $ U(1) $ gauge fluctuations on the same footing and
  find the $ U(1) $ charge fluctuation turns the fixed line into a
  fixed point separating the d-wave superconductor and some unknown
  charge ordered state. Both the dynamic Doppler shift term and the
  mutual statistical interaction become irrelevant short-ranged interactions on both sides of the QCP.
 The spinon and holons are confined into electrons and Cooper pairs in the condensation of $ hc/2e $
  vortices, in contrast to the condensation of double strength $ hc/e $
  vortices discussed in \cite{balents,own} and reviewed in most general form in
  appendix B. There are {\em no} gapless gauge field fluctuations in the final
  theory, in contrast to $ U(1) $ and $ SU(2) $ gauge theory \cite{u1}.
 We also discuss the properties of the two stable phases: disorder
 and superconducting phases around the critical point.

    As stressed in Ref.\cite{static}, in weakly type-II limit $ \xi < \lambda < d_{v} $,
   the superfluid velocity vanishes in the interior of the superconductor, the Doppler shift
   effect is completely absent, the AB phase scattering effect becomes the sole scattering
   mechanism \cite{typeone}. In this case, the interaction between vortices become short-ranged.
   The discussions in \cite{static} on high $T_{c} $ cuprates are limited to $ H_{c1} \ll H \ll H_{c2} $
   where $ \xi \ll d_{v} \ll \lambda $. Except $ H $ is extremely close to $ H_{c1} $, we can safely
   take $ \lambda \rightarrow \infty $ limit. The interaction between vortices become long-ranged
   logarithmic interaction and played leading roles as demonstrated in \cite{static}. In this paper,
   the quantum or thermal fluctuation generated free vortex density
   scales as $ n_{f} \sim \xi^{-2}_{c} $ ( here $ \xi_{c} $ is the correlation length
   rather than the coherence length $ \xi $).
   For high $ T_{c} $ cuprates, $ \kappa= \lambda/\xi \sim 60 $, $ \xi/a \sim 5 $,
   then $ \lambda \sim 60 \xi \sim 300 a $,
   so as long as $ d_{v} \sim \xi_{c} < \lambda $, namely except we get extremely close to the phase boundary,
   we can safely set $ \lambda \rightarrow \infty $. Then we have to consider the long-range interaction between
   the vortices which is mediated by charge gauge field, its important role will be demonstrated
   explicitly in Sec. IV. All the present experimental
   situations lie in the regime $ \xi_{c} < \lambda $.

   The paper is organized as the following. In the next section, we derive
 the effective action with both the mutual C-S interaction and the
 charge fluctuation, first in the path integral language, and then in
 canonical quantization representation. We stress the importance
 of the dual singular gauge transformation on the moving vortices to
 keep the mutual statistics invariant and derive the important
 expression of {\em real} electron operator in terms of spinon and holon (Eqn.\ref{tough}).
 In Sec. III, neglecting the charge fluctuation,
 we concentrate on the effect of the mutual C-S interaction
 and calculate the critical exponents and universal conductivities for
 general statistical angle $ \theta $. Implications for double layers
 Quantum Hall System are given. Some technical details are relegated to appendix A.
 In appendix B, we study double strength $ hc/e $ vortices in the most
  general form by assuming moving $ hc/e $ vortices also scatter
  quasi-particles by a mutual C-S term with any {\em integer} statistical angle.
 In Sec IV, we discuss the combined effect of mutual C-S interaction and charge
 fluctuation, we show how the long-range charge fluctuation change
 the physical picture in Sec. III dramatically. There are only
 electrons and Cooper pairs in the spectrum; the fixed line is replaced by
 a quantum critical point (QCP) separating d-wave superconductor
 and some still unknown charge -ordered state;
 there are no spin-charge separation and no gapless gauge field fluctuations in the final theory, in 
 contrast to $ U(1) $ or $ S(2) $ gauge theories investigated in Ref.\cite{u1}.
 We also discuss the properties of the two phases around the QCP. In section V,
 we apply the formalism developed for quantum generated vortices to study
 thermally generated vortices and random static vortex array in an external magnetic field
  inside $d$-wave superconductor and recovered the results established previously in Ref.\cite{static}.
 Finally, in Sec. VI, we discuss the connection and differences of our present approach
 to the $ Z_{2} $ gauge theory and point out some open problems.

\section{ The dual singular gauge transformation, effective action and
     single electron operator }
    In the following, we follow the notation in Ref.\cite{and,static}
  and introduce the spinon by performing a general single-valued Anderson singular
  gauge transformation $ d=Ud_{s} $:
\begin{equation}
    H_{s}=U^{-1} H U,~~~~~~~~U=
     \left ( \begin{array}{cc}
		e^{i \phi_{A}}  &   0  \\
		0  &   e^{-i \phi_{B} }  \\
		\end{array}   \right )
\label{first}
\end{equation}
 where $ \phi_{A}=0, \phi_{B}= \phi $ or vice versa and $ \phi $ is the phase of the Cooper pair.
 In the former(latter), the spinon is electron-like ( hole-like).
 The original Anderson singular gauge transformation \cite{and} is devised for static vortices.
 Here we extend it to moving vortices whose phase $ \phi $
 is also fluctuating, therefore, depends on both the space and time.

\subsection{ Path integral formulation}
   Expanding $ H_{s} $ around the node 1 where $ \vec{p}=(p_{F},0) $,
   we obtain $ H_{s}=H_{l} + H_{c} $.
   The corresponding linearized quasi-particle Lagrangian $ {\cal L}_{qp} $
   in the presence of the external gauge potential $ A_{\mu} $
   in {\em the imaginary time $ \tau $} is:
\begin{eqnarray}
   {\cal L}^{u}_{qp} &= & \psi^{\dagger}_{1} [ (\partial_{\tau} -i a_{\tau} )
    + v_{f} (p_{x}-  a_{x} ) \tau^{3} + v_{\Delta} (p_{y}-  a_{y} ) \tau^{1} ]
    \psi_{1}    \nonumber   \\
  & + & \psi^{\dagger}_{1} \psi_{1} v_{f} v_{x}(\vec{r}) 
    + i \psi^{\dagger}_{1} \tau^{3} \psi_{1} v_{\tau}(\vec{r}) 
    + (1 \rightarrow 2, x \rightarrow y )
\label{linear}
\end{eqnarray}
   where $ v_{f}, v_{\Delta} $ are Fermi and gap velocities respectively,
   $ v_{\mu}= \frac{\hbar}{2} \partial_{\mu} \phi-\frac{e}{c} 
   A_{\mu}, \mu=x, y, \tau $ is the {\em dynamical} gauge-invariant
   superfluid {\em momentum}, it acts
   as a scalar scattering potential and provides {\em dynamical} Doppler shift
   to the quasi-particles (this effect can be considered as a dynamic Volovik effect),
   $ a_{\mu}= \frac{1}{2} \partial_{\mu} ( \phi_{A} -\phi_{B} ) $ is the {\em dynamical} AB
   gauge field due to the phase winding of vortices.
  Note that the external gauge potential only appear explicitly in the
  superfluid momentum $ v_{\mu} $. The fermion at node 1 (2) only couples to
  the $ x (y) $-component of superfluid momentum, because the Fermi
  momentum at node 1 (2) is along the $x (y) $ direction.
   Because there are equal number of positive and negative vortices,
   on the average, the vanishing of $ v_{\mu} $ and $ a_{\mu} $ is
   automatically ensured in the Anderson gauge.

    We get the corresponding expression at node $ \bar{1} $ and $ \bar{2} $
   by changing
   $ v_{f} \rightarrow -v_{f}, v_{\Delta} \rightarrow -v_{\Delta} $
   in the above Eqn.
\begin{eqnarray}
   {\cal L}^{l}_{qp} &= & \psi^{\dagger}_{\bar{1}} [ (\partial_{\tau} -i a_{\tau} )
    - v_{f} (p_{x}-  a_{x} ) \tau^{3} - v_{\Delta} (p_{y}-  a_{y} ) \tau^{1} ]
    \psi_{\bar{1}}    \nonumber   \\
  & - & \psi^{\dagger}_{\bar{1}} \psi_{\bar{1}} v_{f} v_{x}(\vec{r}) 
    + i \psi^{\dagger}_{\bar{1}} \tau^{3} \psi_{\bar{1}} v_{\tau}(\vec{r}) 
    + (\bar{1} \rightarrow \bar{2}, x \rightarrow y )
\end{eqnarray}

    The curvature term $ H_{c} $ can be written as:
\begin{equation}
   H_{c} =  \frac{1}{m} [  \{ \Pi_{\alpha}, v_{\alpha} \}
	  +  \frac{  \vec{\Pi}^{2}+ \vec{v}^{2} }{2} \tau^{3} 
	  + \frac{\Delta_{0}}{2 \epsilon_{F}} \{ \Pi_{x}, \Pi_{y}\} \tau^{1} ]
\label{curv}
\end{equation}
     Where $ \vec{\Pi}= \vec{p}+  \vec{a} $ is the covariant
     derivative \cite{wrong}.
     $ H_{c} $ takes the {\em same } form for all the four nodes.
     Although $ H_{c} $ is very important for the thermal Hall
    conductance in the presence of external magnetic field
    \cite{static}, it is irrelevant near the quantum critical point,
     so we will not discuss it anymore in the present paper.

   Performing a $ P-H $ transformation
   $ \tilde{\psi}_{1\alpha}= \epsilon_{\alpha \beta}
   \psi^{\dagger}_{\bar{1}\beta} $ ( $\alpha, \beta $ are p-h indices ) at node $ \bar{1} $
  and the corresponding transformation at node $ \bar{2} $, the above Eqn. becomes:
\begin{eqnarray}
   {\cal L}^{l}_{qp} &= & \tilde{\psi}^{\dagger}_{1} [ (\partial_{\tau} +
   i a_{\tau} )
    + v_{f} (p_{x}+  a_{x} ) \tau^{3} + v_{\Delta} (p_{y}+  a_{y} ) \tau^{1} ]
    \tilde{\psi}_{1}    \nonumber   \\
  & + & \tilde{\psi}^{\dagger}_{1} \tilde{\psi}_{1} v_{f} v_{x}(\vec{r}) 
  + i \tilde{\psi}^{\dagger}_{1} \tau^{3} \tilde{\psi}_{1} v_{\tau}(\vec{r}) 
  + (1 \rightarrow 2, x \rightarrow y )
\label{tilde}
\end{eqnarray}

   In order to make the final expressions explicitly $ SU(2) $ invariant,
   we perform the singular gauge transformation 
   $ \psi_{12\alpha}= e^{-i(\phi_{A}-\phi_{B} )}
    \tilde{\psi}_{1 \alpha} $ and the
  corresponding transformation at node $ \bar{2} $, then $ a_{\mu}
  \rightarrow -a_{\mu} $, Eqn.\ref{tilde} takes
  the same form as Eqn.\ref{linear} \cite{break}. Adding the two equations
   leads to: 
\begin{eqnarray}
   {\cal L}_{qp} &= & \psi^{\dagger}_{1a} [ (\partial_{\tau} -i a_{\tau} )
    + v_{f} (p_{x}-  a_{x} ) \tau^{3} + v_{\Delta} (p_{y}-  a_{y} ) \tau^{1} ]
    \psi_{1a}    \nonumber   \\
  & + & \psi^{\dagger}_{1a} \psi_{1a} v_{f} v_{x}(\vec{r}) 
    + i \psi^{\dagger}_{1a} \tau^{3} \psi_{1a} v_{\tau}(\vec{r}) 
    + (1 \rightarrow 2, x \rightarrow y )
\label{linear4}
\end{eqnarray}
  where $ a=1,2 $ is the spin indices. $ \tau^{\prime s} $ matrices are acting
  on {\em particle-hole} space.
   As intended, the above Eqn. is explicitly $ SU(2) $ spin invariant.
  Equivalently, we can start with the explicitly spin $ SU(2) $ 
  invariant approach advocated in Ref.\cite{balents} and perform the 
  singular gauge transformation in the {\em p-h } space. 

 $ {\cal L}_{qp} $ enjoys gauge symmetry $ U(1)_{u}
  \times U_{s}(1) $ ( in fact, for static vortices,
   it is $ U(1) \times Z_{2} $ ),
  the first being uniform and second being staggered 
  gauge symmetry:
\underline{Uniform (or external) $ U_{u}(1) $ gauge symmetry }
\begin{eqnarray}
  c_{\alpha} & \rightarrow & c_{\alpha} e^{i \chi},~~~ 
  d_{\alpha} \rightarrow d_{\alpha}   \nonumber \\
  \phi_{A} & \rightarrow & \phi_{A} + \chi ,~~~~
  \phi_{B} \rightarrow \phi_{B} + \chi
\label{ext}
\end{eqnarray}
   Under this uniform $ U(1) $ transformation, the corresponding fields
   transform as:
\begin{eqnarray}
  \phi & \rightarrow & \phi + 2\chi ,~~~~ A_{\alpha} \rightarrow 
  A_{\alpha} + \partial_{\alpha} \chi
	     \nonumber   \\
  v_{\alpha} & \rightarrow & v_{\alpha},~~~~~a_{\alpha} \rightarrow a_{\alpha}
\label{uniform}
\end{eqnarray}

      $ d, v_{\alpha}, a_{\alpha} $ all are invariant under this external
      $ U(1) $ transformation. Therefore the spinon $ d_{\alpha} $
      is charge {\em neutral} to the external magnetic field.

\underline{Staggered (or internal) $ U_{s}(1) $ gauge symmetry}
\begin{eqnarray}
  c_{\alpha} & \rightarrow & c_{\alpha},~~~ 
  d_{\alpha} \rightarrow d_{\alpha} e^{-i \chi}   \nonumber \\
  \phi_{A} & \rightarrow & \phi_{A} + \chi ,~~~~
  \phi_{B} \rightarrow \phi_{B} - \chi
\label{int}
\end{eqnarray}
   Under this internal $ U(1) $ transformation, the corresponding fields
   transform as:
\begin{eqnarray}
  \phi  & \rightarrow & \phi ,~~~~ A_{\alpha} \rightarrow A_{\alpha} 
	     \nonumber   \\
   v_{\alpha} & \rightarrow & v_{\alpha},~~~~~a_{\alpha} \rightarrow a_{\alpha}+
      \partial_{\alpha} \chi
\label{stagger}
\end{eqnarray}

      Although the spinon $ d_{\alpha} $
      is charge neutral to the external magnetic field, it carries
      charge $ 1 $ to the internal gauge field $ a_{\alpha} $.

   It is easy to realize that $ U_{u}(1) $ acts only on the boson sector,
  since the spinon is charge neutral, $ U_{s}(1) $ acts only on the
   fermion sector. In fact, $ U_{s}(1) $ should
   be a discrete local $ Z_{2} $ symmetry for static vortices,
   because up and down static
  $ hc/2e $ vortices are equivalent and do not break T symmetry \cite{static}.

     The phase fluctuation is simply $ 2+1 $ dimensional X-Y model
    (For simplicity, we neglect the possible anisotropy in the spin-stiffness):
\begin{equation}
    {\cal L}_{ph}= \frac{K}{2} v^{2}_{\mu} =
    \frac{K}{2} ( \partial_{\mu} \phi -2 A_{\mu} )^{2}
\end{equation}
  
    After absorbing the scalar potential scattering part (the dynamic Doppler shift term)
   into $ {\cal L}_{ph} $,   we can write the total Lagrangian
   $ {\cal L}= {\cal L}_{qp}+ {\cal L}_{ph} $ as:
\begin{eqnarray}
   {\cal L} &= & \psi^{\dagger}_{1a} [ (\partial_{\tau} -i a_{\tau} )
    + v_{f} (p_{x}-  a_{x} ) \tau^{3} + v_{\Delta} (p_{y}-  a_{y} ) \tau^{1} ]
    \psi_{1a}    \nonumber   \\
    & + &  (1 \rightarrow 2, x \rightarrow y )
      + \frac{K}{2} ( \partial_{\mu} \phi - A^{eff}_{\mu} )^{2}
\label{act}
\end{eqnarray}
   where $ A^{eff}_{\mu}= 2 A_{\mu} -K^{-1} J_{\mu} $ and the quasi-particle
   electric current is : $ J_{0} = \psi^{\dagger}_{j} \tau^{3}
   \psi_{j}, J_{x} = v_{f} \psi^{\dagger}_{1} \psi_{1},
   J_{y} =v_{f} \psi^{\dagger}_{2} \psi_{2} $.  It is easy to see that
   the dynamical Doppler shift term on the quasi-particles can also be equivalently
   thought as a dynamical effective gauge field coupled to the phase fluctuation.
   (Note that in the static vortex array generated by external magnetic field \cite{static},
   the superfluid momentum is static and causes Doppler shift to the quasi-particle,
   there is no feed-back effect on the superfluid momentum from the quasi-particles) 

    From Eqn.\ref{act}, it is easy to identify the two conserved Noether
    currents: spinon current and electric current.

    The spinon current is given by:
\begin{eqnarray}
   j^{s}_{0} & = & \psi^{\dagger}_{1}(x) \psi_{1}(x) + \psi^{\dagger}_{2}(x) 
                \psi_{2}(x)
                       \nonumber  \\
   j^{s}_{x} & = & \psi^{\dagger}_{1}(x)  v_{F} \tau^{3} \psi_{1}(x)
  + \psi^{\dagger}_{2}(x)  v_{\Delta} \tau^{1} \psi_{2}(x)
                       \nonumber  \\
  j^{s}_{y} & = & \psi^{\dagger}_{1}(x)  v_{\Delta} \tau^{1} \psi_{1}(x)
   + \psi^{\dagger}_{2}(x)  v_{F} \tau^{3} \psi_{2}(x)
\label{gap}
\end{eqnarray}

    Obviously the spinon current only comes from quasi-particle.
  In principle, the spinon current is not conserved due to scatterings between
  different nodes which lead to anomalous terms not included in Eqn.\ref{act}.
  However, the inter-node scatterings involve large momenta transfer
  $ \vec{K}_{i} -\vec{K}_{j} $ for $ i \neq j $,
  so we neglect them due to momentum conservation
 in the long wave-length limit of the phase fluctuation \cite{magnetic}.
   The exact conserved {\em spin current}
  $ \vec{j}^{S}_{\mu} $ is with $ \vec{\sigma}/2 $ inserted in the above
  spinon currents.

     The electric current is given by:
\begin{equation}
    j^{e}_{\mu}= -\frac{\partial {\cal L}}{ \partial A^{eff}_{\mu} }
                  = K (\partial_{\mu} \phi - A^{eff}_{\mu})
                  = K (\partial_{\mu} \phi - 2 A_{\mu}) + J_{\mu}
\end{equation}
 Where the first part coming from Cooper pair and the second from
 the quasi-particle. Although they are not separately conserved, their sum is.
    
    In the the electron-like Anderson gauge where $ a_{\mu}= \frac{1}{2} \partial_{\mu} \phi $
    \cite{hole}, the Noether current due to the the symmetry under $ \phi \rightarrow \phi
   + \chi $ can be written as:
\begin{equation}
    j^{t}_{\mu}= \frac{\partial {\cal L}}{ \partial (\partial_{\mu} \phi)}
                  = K (\partial_{\mu} \phi - A^{eff}_{\mu}) 
                    - \frac{1}{2} j^{s}_{\mu}
                  =  j^{e}_{\mu} -  \frac{1}{2} j^{s}_{\mu}
\end{equation}
    
     It is a combination of electric and spinon currents, therefore also
   conserved.

    Following Refs.\cite{dual,berry}, we perform a duality transformation
   to Eqn.\ref{act}
\begin{eqnarray}
   & &    \frac{K}{2} ( \partial_{\mu} \phi - A^{eff}_{\mu} )^{2} 
   -i \frac{1}{2} \partial_{\mu} \phi j^{s}_{\mu}
            \nonumber  \\
   &= & i j_{e \mu} ( \partial_{\mu} \phi - A^{eff}_{\mu} ) + \frac{1}{2 K}
     j^{2}_{ e \mu} -i \frac{1}{2} \partial_{\mu} \phi j^{s}_{\mu}
                      \nonumber  \\
   &= & i j^{t}_{\mu} \partial_{\mu} \theta + 
    i  j^{t}_{ \mu} \partial_{\mu} \phi 
    - i j^{e}_{\mu} A^{eff}_{\mu}  +  \frac{1}{2 K} j^{2}_{ e \mu}
\label{boson}
\end{eqnarray}
   Where we have separated topological trivial spin-wave part and
   topological non-trivial vortex parts.

    Integrating out the spin-wave part, we get the conservation equation for
  the total current $ j^{t}_{\mu}= j^{e}_{\mu} - 
  \frac{1}{2} j^{s}_{\mu} $. In fact, as
  shown in the previous paragraphs, $ j^{e}_{\mu} $ and $ j^{s}_{\mu} $ are
  separately conserved. Therefore we can introduce spin and electric gauge
  fields by:
\begin{eqnarray}
     j^{s}_{\mu} & = & \epsilon_{\mu \nu \lambda } \partial_{\nu} a^{s}_{\lambda}
               \nonumber  \\
     j^{e}_{\mu} &= & \epsilon_{\mu \nu \lambda } \partial_{\nu} a^{e}_{\lambda}
\end{eqnarray}

      We can also define the vortex current:
\begin{eqnarray}
     j^{v}_{\mu} & = & \frac{1}{2 \pi} 
    \epsilon_{\mu \nu \lambda } \partial_{\nu} \partial_{\lambda} \phi 
                \nonumber  \\
     & = & \epsilon_{\mu \nu \lambda } \partial_{\nu} a^{v}_{\lambda}
\end{eqnarray}
      where $ a^{v}_{\mu} = \partial_{\mu} \phi/ 2 \pi $ is the vortex gauge
   field.
   
     Substituting the above expressions into Eqn.\ref{boson}, we reach:
\begin{eqnarray}
  & &  \frac{1}{ 4 K} f^{2}_{e \mu \nu} + i \partial_{\mu} \phi j^{t}_{\mu}
   -i A^{eff}_{\mu} \epsilon_{\mu \nu \lambda } \partial_{\nu} a^{e}_{\lambda}
            \nonumber  \\
  & = & \frac{1}{ 4 K} f^{2}_{e \mu \nu} + i a^{t}_{\mu} j^{v}_{\mu}
   -i A^{eff}_{\mu} \epsilon_{\mu \nu \lambda } \partial_{\nu} a^{e}_{\lambda}
\end{eqnarray}
    where $ a^{t}_{\mu}= a^{e}_{\mu}- \frac{1}{2} a^{s}_{\mu} $ is the total
  gauge field felt by the moving vortices.

    Using $ \Phi $ for the vortex operator and adding the 
  quasi-particle part, we get the following effective action:
\begin{eqnarray}
   {\cal L} &= & \psi^{\dagger}_{1a} [ (\partial_{\tau} -i a^{\psi}_{\tau} )
    + v_{f} (p_{x}-  a^{\psi}_{x} ) \tau^{3} + v_{\Delta} (p_{y}-  a^{\psi}_{y} )
    \tau^{1} ]
    \psi_{1a}   \nonumber  \\
       &  + &  (1 \rightarrow 2, x \rightarrow y )  
                       \nonumber  \\
  & + & | ( \partial_{\mu} -i a^{\Phi}_{\mu}-i a_{\mu} ) \Phi |^{2}
      + V( |\Phi|) + \frac{i}{ 2 \pi \theta} a^{\psi}_{\mu} 
    \epsilon_{\mu \nu \lambda } \partial_{\nu} a^{\Phi}_{\lambda}
                        \nonumber  \\
  & + & \frac{1}{ 4 } f^{2}_{ \mu \nu} 
   -i A^{eff}_{\mu} \epsilon_{\mu \nu \lambda } \partial_{\nu} a_{\lambda}
    - \mu \epsilon_{ij} \partial_{i} a_{j}
\label{dual}
\end{eqnarray}
  where $ V(|\Phi|)= m^{2}_{\Phi} |\Phi|^{2}+ g_{\Phi} |\Phi|^{4} +\cdots $
  stands for the short range interaction between the vortices.
  The last term is due to the Berry phase in the boson representation
  \cite{berry} which is a first order time derivative term.
   It can be absorbed into $ A^{eff}_{\mu} $ by redefining 
   $ A^{eff}_{\mu} \rightarrow A^{eff}_{\mu} +i \mu \delta_{\mu 0} $,
   it acts like an external magnetic field in the $ \hat{z} $
  ( namely $\mu=0 $ ) direction in the above vortex representation.
  There are $ M=1 $ species of vortex and $ N=4 $ species of Dirac
  fermion( 2 spin components at 2 nodes at the upper half plane),
  the mutual statistical angle $ \theta= \pm 1/2 $. We also changed
  the notation by setting $ a^{\psi} = a^{v}, a^{\Phi}= a^{s}, a^{e}=a $.
     The dynamic Doppler-shifted term
     $ i K^{-1} J_{\mu} \epsilon_{\mu \nu \lambda} \partial_{\nu} a_{\lambda} $
     (The dynamic Doppler shift effect) is encoded in the second to the last term
     in Eqn.\ref{dual}. Note that this term only involves electrical gauge field.
      This effect was not written down explicitly in
     the $ Z_{2} $ gauge theory \cite{z2}.

    Eqn.\ref{dual} again enjoys the gauge symmetry $ U(1)_{c} \times U(1)_{cs} $ which is a dynamic
    generalization of $ U(1)_{u} \times U(1)_{s} $ in Eqns.\ref{uniform},\ref{stagger}.
   The first one acting on the vortex sector is the electric $ U(1)_{c} \sim U(1)_{u} $
   gauge field, the second on both
   the vortex and fermion sectors is the $ U(1)_{cs} $ mutual CS gauge field which is a dynamic
    generalization of $ U(1)_{s} $.
    The mutual Chern-Simon term enforces the constraints:
   $ 2 \pi \theta j^{s}_{\mu}  =  \epsilon_{\mu \nu \lambda } \partial_{\nu} 
    a^{\Phi}_{\lambda}, 2 \pi \theta j^{v}_{\mu}  =  \epsilon_{\mu \nu \lambda }
    \partial_{\nu} a^{\psi}_{\lambda} $. Physically, it means that when 
    a quasi-particle encircles
    a vortex, it picks up a phase $ 2 \pi \theta $. Equivalently
    when a vortex moves around a quasi-particle, it also picks up a phase
    $ 2 \pi \theta $.  Although the conventional C-S term breaks T symmetry
    and has periodicity under $ \theta \rightarrow \theta +2 $\cite{zhang,hlr,chen,wen,boson,subir}, the
    mutual C-S term does {\em not } break T symmetry and does {\em not}
     have the periodicity under $ \theta \rightarrow \theta +1 $. For example,
     $ \theta=-1/2 $ is {\em not} equivalent to $ \theta=1/2 $. 

\subsection{Electron operator}
    The general form of electron annihilation operator is:
\begin{equation}
   C_{\alpha}(\vec{x})= \sum_{i=1,2} [ e^{i \phi_{A} }
    e^{i \vec{K}_{i} \cdot \vec{x}}
    \psi_{i 1 \alpha} -\epsilon_{\alpha \beta}
    e^{i \phi_{B} } e^{-i \vec{K}_{i} \cdot \vec{x}}
    \psi^{\dagger}_{i 2 \beta} ]
\label{gen1}
\end{equation}
   where $ i $ is the node index, $1, 2 $ are $ p-h $ indices,
 $ \alpha, \beta $ are spin indices (in the rest of the paper,
 we will stick to this notation ) and $ \phi_{A} $ and $ \phi_{B} $ 
 are given by:
\begin{eqnarray}
  \phi_{A} & = & \frac{\phi}{2} + \int a^{\psi} d x  \nonumber  \\
  \phi_{B} & = & \frac{\phi}{2} - \int a^{\psi} d x
\label{ab}
\end{eqnarray}

    Eqn.\ref{gen1} reads:
\begin{eqnarray}
   C_{\alpha}(\vec{x}) & = & \sum_{i=1,2} [ e^{i \phi/2 }
    e^{i \vec{K}_{i} \cdot \vec{x}} e^{i \int a^{\psi} d x} \psi_{i 1 \alpha}
                  \nonumber  \\
  & - & \epsilon_{\alpha \beta}
    e^{i \phi/2 } e^{-i \vec{K}_{i} \cdot \vec{x}}
    e^{-i \int a^{\psi} d x} \psi^{\dagger}_{i 2 \beta} ]
\label{tough}
\end{eqnarray}
    
   In principle, the electron Green function $ G(\vec{x},t)=
   < C_{\alpha}(\vec{x},t) C^{\dagger}_{\alpha}(0,0) > $ can be calculated
   from the above equation. It is easy to see that
   the real electron in Eqn.\ref{tough} is invariant under the
   internal gauge transformation $U(1)_{s}$.

\subsection{Canonical quantization approach}

   In subsection A, we derived Eqn.\ref{dual} by path integral approach, in this subsection, we rederive
  this equation by canonical quantization method and stress the importance of the dual singular 
  gauge transformation Eqn.\ref{mcs2}.

  Substituting Eqn.\ref{ab} into Eqn.\ref{first}, we find that
 the singular unitary transformation $ U $ takes the physically more
 transparent form:
\begin{equation}
    d = \left ( \begin{array}{cc}
		e^{i \phi/2 }  &   0  \\
		0  &   e^{-i \phi/2 }  \\
		\end{array}   \right )
             e^{i \int a^{\psi} dx} d_{s}
\label{mcs1}
\end{equation}
  where, in {\em first } quantization form:
\begin{equation}
  \vec{a}^{\psi}= \frac{1}{2} ( \nabla \phi_{A} - \nabla \phi_{B} )
\end{equation}

  The first factor leads to the dynamic Doppler shift term which is not explicitly written in the
  $ Z_{2} $ gauge theory. The second leads to mutual CS term with $ \theta=1/2 $.
  To see this, in Anderson electron-like gauge, we rewrite $ \vec{a^{\psi}} $
  in the second quantization form in terms of the vortex operator $ \Phi $ :
\begin{equation}
    \vec{a}^{\psi}=  \theta  \int d^{2} \vec{r}^{\prime}
   \nabla \phi(\vec{r}-\vec{r}^{\prime}) j^{v}_{0} (\vec{r}^{\prime})
\label{con1}
\end{equation}
  where $ j^{v}_{0} $ is vortex number operator and
  $ \phi( \vec{r}-\vec{r}^{\prime} ) $ is the angle the vector
  $ (\vec{r}-\vec{r}^{\prime} ) $ makes with $ x $ axis.

  Obviously, $ d_{s} $ still satisfies fermion commutation relation. When the vortices are static, $ j^{v}_{0}(\vec{x})
  =\sum_{i} \delta( \vec{x}-\vec{x}_{i} ) $, Eqn.\ref{con1} recovers the original Anderson singular gauge transformation
  and is the only necessary transformation. However, when vortices are generated by quantum fluctuations,
  they have their {\em own dynamics}.
  Eqn.\ref{con1} alone can {\em not} satisfy all the possible commutation relations. This motivated us to
  introduce the following {\em dual} singular gauge
  transformation to the vortex operator:
\begin{equation}
    \Phi = e^{i \int a^{\Phi} dx} \Phi_{s}
\label{mcs2}
\end{equation}
   where, in {\em second} quantization form:
\begin{equation}
    \vec{a}^{\Phi}=  \theta  \int d^{2} \vec{r}^{\prime}
   \nabla \phi(\vec{r}-\vec{r}^{\prime}) j^{s}_{0}(\vec{r}^{\prime})
\label{con2}
\end{equation}
  Where $ j^{s}_{0} $ is given by Eqn.\ref{gap}.

  Obviously, $ \Phi_{s} $ still satisfies boson commutation relation.

  Most importantly, from Eqns.\ref{mcs1} and \ref{mcs2},
  we can check that $ d_{s} $ and $ \Phi_{s} $
  commute with each other, so their mutual statistics is kept intact which is our original motivation.
  Note that for static vortices, the dual transformation in Eqn.\ref{mcs2}
  is unnecessary, because the static vortices are not dynamic variables.

     It is very instructive to compare the above transformations to the
  well-known singular gauge transformation leading to composite boson
   in $ \nu=1/3 $ Fractional Quantum Hall state 
   by Zhang, Hansen and Kivelson (ZHK) \cite{zhang}
   and composite fermion
  in $ \nu=1/2 $ system by Halperin, Lee and Read (HLR) \cite{hlr}:
\begin{equation} 
    \psi= e^{i \int a dx} \psi_{c}
\label{ccs}
\end{equation}
   where, in {\em second} quantization form:
\begin{equation}
    \vec{a}=  \theta  \int d^{2} \vec{r}^{\prime}
   \nabla \phi(\vec{r}-\vec{r}^{\prime}) \rho(\vec{r}^{\prime})
\label{con3}
\end{equation}
  where $ \rho( \vec{r} ) = \psi^{\dagger} \psi =\psi^{\dagger}_{c} \psi_{c} $ is the fermion number operator.

  The crucial difference between Eqn.\ref{ccs}
  and Eqns.\ref{mcs1},\ref{mcs2} is that in the former,
  we attach electron's own $ \theta $
  flux to itself ( $ \theta=3 $ in $ \nu=1/3 $ FQH state to change fermion to
  boson and
 $ \theta =2 $ in $ \nu=1/2 $ system to keep fermion
  statistics intact ), the constraint  $ \nabla \times \vec{a}(\vec{r})
  = 2 \pi \theta \rho(\vec{r}) $ ( Eqn.\ref{con3} ) in Coulomb gauge
  $ \partial_{i} a_{i} =0 $
  leads to composite fermion $ \psi_{c} $ coupled
  to conventional CS term which breaks T symmetry, has periodicity
  under $ \theta \rightarrow \theta +2 $; however,
  in the latter, we attach vortex's
  $ \theta=1/2 $ flux to quasi-particle or {\em vise versa},
  the two constraints $ \nabla \times \vec{a}^{\psi}(\vec{r})
  = 2 \pi \theta j^{v}_{0}(\vec{r}), \nabla \times \vec{a}^{\Phi}(\vec{r})
  = 2 \pi \theta j^{s}_{0}(\vec{r}) $
  ( Eqns.\ref{con1},\ref{con2} ) in Coulomb gauges
  $ \partial_{i} a^{\psi}_{i} = \partial_{i} a^{\Phi}_{i} =0 $
  naturally leads to quasi-particles and vortices are
  coupled by mutual CS term in additional to the electrical gauge
  field which mediates the logarithmic interaction between vortices.
  The mutual CS term does not break T symmetry and does not have the
  periodicity under $ \theta \rightarrow \theta +1 $.

   It is easy to see that when a vortex is moving around a 
  closed loop, it pick up two phases, one is $ U(1)_{cs} $ phase
  due to the spinon current, described
  by the mutual C-S term, another is $ U(1)_{c} $ phase
  due to the total electric charge current
  described by the Maxwell term. In the following section, we will neglect
  the electric charge fluctuation (therefore its associated Dynamic Doppler shift term)
  and concentrate on the mutual statistics term.
  In Sec. IV, we will study the combined effect of charge fluctuation described by the
  Maxwell term and the mutual C-S term.
   
     For simplicity, we neglect all the possible anisotropy in the quasi-particle and vortex velocities.
    We also take the relativistic form for both fermion and boson,
   because the difference between the velocity of spinon and that of vortex
   in Eqn.\ref{dual} is expected to be irrelevant
   near the zero temperature QCP \cite{guess}.

\section{ The effect of mutual statistics }
    In this section, we neglect the charge fluctuation, namely setting
    $ a_{\mu}=0 $ in Eqn.\ref{dual}. The associated dynamic Doppler shift term also drops out.
    As pointed out in \cite{static} and re-emphasized in the introduction, this is
    the weakly type II case. So far, weakly type II superconductors are all s-wave superconductors where
    there is no low energy quasi-particles. So the results in this section can only
    be applied to presumed " weakly type II d-wave superconductor " which has not been
    discovered yet in nature. The model is also a relativistic version of the Hall drag problem
    in double layer Quantum Hall system \cite{kun}, therefore  
    the model itself is interesting on its own right and deserves detailed
    investigation. 

    The charge fluctuation can be suppressed by assuming the condensation of $ hc/e $ vortices
    reviewed in appendix B.
    We could add the kinetic and potential terms for the $ hc/e $ vortex
    operator $ \Phi_{2} $ to Eqn.\ref{dual}:
\begin{equation}
  {\cal L}_{\Phi_{2}}= | ( \partial_{\mu} -i 2 a_{\mu} ) \Phi_{2} |^{2}
   + V( |\Phi_{2}|) 
\end{equation}
     As shown in \cite{balents} and in appendix B, there is no mutual statistical interaction
   between spinon and $ hc/e $ vortices $ \Phi_{2} $. The long range
   logarithmic interaction between $ hc/2e $ vortices $ \Phi $ and $ hc/e $
   vortices $ \Phi_{2} $ is mediated by the electrical gauge field $ a_{\mu} $.
   Condensing $ < \Phi_{2} > = \Phi_{20} $ will generate a mass term
   $ \frac{\Phi^{2}_{20}}{2} (a_{\mu})_{t}^{2} $ which dominates 
   over the Maxwell term. Integrating out $ a_{\mu} $ leads to
\begin{equation}
   \frac{1}{\Phi^{2}_{20} }[ \frac{1}{ 4} ( f^{A^{eff}}_{\mu \nu} )^{2} 
   -i  A^{eff}_{\mu} \epsilon_{\mu \nu \lambda } \partial_{\nu} 
   j^{v}_{\lambda} + (j^{v}_{\mu})^{2}_{t}  ]
\end{equation}
  
 All the generated terms  only renormalize
   the short range interactions already included in $ V(|\Phi| ) $.
\subsection{ Quantum Critical point}
   In order to calculate conductivities, we add two source fields
   $ A^{\psi}_{\mu} $ and $ A^{\Phi}_{\mu} $ for the quasi-particles and
    vortex respectively:

\begin{eqnarray}
   {\cal L} &= & \psi^{\dagger}_{a} \gamma_{\mu} (\partial_{\mu}
    -i a^{\psi}_{\mu} -i A^{\psi}_{\mu} ) \psi_{a}   \nonumber  \\
  & + & | ( \partial_{\mu} -i a^{\Phi}_{\mu}-i A^{\Phi}_{\mu} ) \Phi |^{2}
      + V( |\Phi|) + \frac{i}{ 2 \pi \theta} a^{\psi}_{\mu} 
    \epsilon_{\mu \nu \lambda } \partial_{\nu} a^{\Phi}_{\lambda}
\label{mutual}
\end{eqnarray}
    where $ a=1,2,3,4 $ stands for $ N=4 $ species of Dirac fermion.
   In fact, two Maxwell terms for $ a^{\psi} $ and $ a^{\Phi} $ can be
   added to the above equation, but they are subleading to the mutual CS
   term in the low energy limit. It is expected that there is no periodicity
   under $ \theta \rightarrow \theta +1 $ in the continuum limit.

    The RG calculation in Refs.\cite{chen,wen,boson} can be used to show
   that $ \theta $
   is exactly marginal, therefore there is a line of fixed points determined
   by the mutual statistical angle $ \theta $.
   In order to calculate the {\em spin conductivity} along this fixed line,
   a source field could be introduced to couple to  the spin current
   $ \vec{j}^{S}_{\mu}=
   \psi^{\dagger}_{\alpha}
    \gamma_{\mu} (\vec{\sigma})_{\alpha \beta} \psi_{\beta} $.
   Similar calculations follow.

     Integrating out both fermion and boson leads to:
\begin{eqnarray}
   {\cal L} & = & -\frac{1}{2} a^{\psi}_{\mu} (-k)
         \Pi^{\psi}_{\mu \nu}(k) a^{\psi}_{\nu}(k)
    -\frac{1}{2} a^{\psi}_{\mu} (-k)
         \Pi^{\psi \Phi}_{\mu \nu}(k) a^{\Phi}_{\nu}(k)          
                     \nonumber  \\
   & - & \frac{1}{2} a^{\Phi}_{\mu} (-k)
         \Pi^{\Phi}_{\mu \nu}(k) a^{\Phi}_{\nu}(k)
    -\frac{1}{2} a^{\Phi}_{\mu} (-k)
         \Pi^{\Phi \psi}_{\mu \nu}(k) a^{\psi}_{\nu}(k)
                      \nonumber  \\
   & - & \frac{1}{2 \theta} ( a^{\psi}_{\mu} (-k) - A^{\psi}_{\mu} (-k)) 
         \epsilon_{\mu \nu \lambda} k_{\lambda} ( a^{\Phi}_{\nu}(k)
         - A^{\Phi}_{\nu}(k))
                         \nonumber  \\
   & - &  \frac{1}{2 \theta} ( a^{\Phi}_{\mu} (-k) - A^{\Phi}_{\mu} (-k)) 
         \epsilon_{\mu \nu \lambda} k_{\lambda} ( a^{\psi}_{\nu}(k)
         - A^{\psi}_{\nu}(k))
\label{gen}
\end{eqnarray}
    where the exact forms of $ \Pi^{\prime s} $ are dictated by gauge
   invariance and Furry's theorem :
\begin{eqnarray}
 \Pi^{\psi}_{\mu \nu}(k) &= & \Pi_{1}(k) k ( \delta_{\mu \nu} -
    \frac{ k_{\mu} k_{\nu} }{k^{2}} )   \nonumber  \\
 \Pi^{\psi \Phi}_{\mu \nu}(k) &= & \Pi^{\Phi \psi}_{\mu \nu}(k) = 
 \Pi_{2}(k) \epsilon_{\mu \nu \lambda} k_{\lambda} 
                       \nonumber  \\
 \Pi^{\Phi}_{\mu \nu}(k) &= & \Pi_{3}(k) k ( \delta_{\mu \nu} -
    \frac{ k_{\mu} k_{\nu} }{k^{2}} )
\end{eqnarray}

   Where $ \Pi_{1}, \Pi_{2}, \Pi_{3} $ are the polarizations for
  $ a^{\psi} a^{\psi}, a^{\psi} a^{\Phi}, a^{\Phi} a^{\Phi} $.

  If we are only interested in DC conductivities, for simplicity, we can
  put $ \vec{k} =0 $, Eqn.\ref{gen} becomes ( for the most general form,
    see appendix A): 
\begin{eqnarray}
   {\cal L} & = & -\frac{1}{2} a^{\psi}_{i} (-\omega_{n})
         \Pi_{1} | \omega_{n} | a^{\psi}_{i}(\omega_{n})
    -\frac{1}{2} a^{\psi}_{i} (-\omega_{n})
         \Pi_{2} \epsilon_{ij} \omega_{n} a^{\Phi}_{j}(\omega_{n})          
                     \nonumber  \\
    & - & \frac{1}{2} a^{\Phi}_{i} (-\omega_{n})
         \Pi_{3} | \omega_{n} | a^{\Phi}_{i}(\omega_{n})
    -\frac{1}{2} a^{\Phi}_{i} (-\omega_{n})
         \Pi_{2} \epsilon_{ij} \omega_{n} a^{\psi}_{j}(\omega_{n})          
                      \nonumber  \\
   & - & \frac{1}{2 \theta} ( a^{\psi}_{i} (-\omega_{n}) 
    - A^{\psi}_{i} (-\omega_{n})) 
         \epsilon_{ij} \omega_{n} ( a^{\Phi}_{j}(\omega_{n})
         - A^{\Phi}_{j}(\omega_{n}))
                         \nonumber  \\
   & - & \frac{1}{2 \theta} ( a^{\Phi}_{i} (-\omega_{n}) 
    - A^{\Phi}_{i} (-\omega_{n})) 
         \epsilon_{ij} \omega_{n} ( a^{\psi}_{j}(\omega_{n})
         - A^{\psi}_{j}(\omega_{n}))
\end{eqnarray}

    If we define $ \tilde{a}^{\Phi}_{i} (\omega_{n})
    =\epsilon_{ij} a^{\Phi}_{j} (\omega_{n}),
     \tilde{A}^{\Phi}_{i} (\omega_{n}) 
    =\epsilon_{ij} A^{\Phi}_{j} (\omega_{n}) $,
     the above
    equation becomes diagonal in the spatial indices $ i=1,2 $.
    Finally, integrating out $ a^{\psi}_{i}, \tilde{a}^{\Phi}_{i} $ leads to
\begin{equation}
     {\cal L} = -\frac{1}{2} ( A^{\psi}_{i}, \tilde{A}^{\Phi}_{i} )
                                \left( \begin{array}{cc}
				\sigma^{\psi} &   \sigma^{H}  \\
				\sigma^{H}  &   \sigma^{\Phi}  \\
				\end{array}   \right )
                                \left( \begin{array}{c}
				A^{\psi}_{i}   \\
				\tilde{A}^{\Phi}_{i}  \\
				\end{array}   \right )
\label{matrix}
\end{equation}
   
   Where spinon, vortex and mutual Hall drag conductivities are:
\begin{eqnarray}
   \sigma^{\psi} & = & ( \frac{1}{\theta} )^{2} \frac{\Pi_{1}}{
    \Pi_{1} \Pi_{3} + (1/\theta -\Pi_{2})^{2} }  \nonumber  \\
   \sigma^{\Phi} & = & ( \frac{1}{\theta} )^{2} \frac{\Pi_{3}}{
    \Pi_{1} \Pi_{3} + (1/\theta -\Pi_{2})^{2} }  \nonumber  \\
   \sigma^{H} & = & ( \frac{1}{\theta} )^{2} \frac{\Pi_{2} -\theta ( \Pi_{1}
     \Pi_{3} + \Pi^{2}_{2} )}{
    \Pi_{1} \Pi_{3} + (1/\theta -\Pi_{2})^{2} }
\label{three}
\end{eqnarray}

      In fact, all the three conductivities can be written in the elegant
    connection formula:
\begin{equation}
   \rho_{ij}= (\rho_{FB} )_{ij} - \theta \epsilon_{ij}
\label{conn}
\end{equation}
     with the conductivity tensor of fermion and boson given by
\begin{equation}
                      \sigma_{FB}=  \left( \begin{array}{cc}
				\Pi_{1} &  - \Pi_{2}  \\
				\Pi_{2}  &   \Pi_{3}  \\
				\end{array}   \right )
\label{exact}
\end{equation}

  Although its form is similar to
 the conventional connection formulas discussed in \cite{hlr,chen,wen},
 the physical interpretations of the conductivities are quite different
 (see the following).

   When the vortices are generated by external magnetic field and pinned
  by impurities as discussed in \cite{static}, the total conductivity is
  the same as the fermion conductivity because the static vortices do not
  contribute. As explained in section V, the static vortices only feel
  the electric gauge field $ a_{\mu} $, but {\em not}
  the statistical gauge field $ a^{\Phi}_{\mu} $.

 The equations \ref{conn},\ref{exact} are exact, but $ \Pi_{1}, \Pi_{2}, \Pi_{3} $
 can only be calculated perturbatively in the coupling constant
  $ g^{2}= 2 \pi \theta $. The renormalized propagators for $ a^{\psi} $ and $ a^{\Phi} $
  can be found from the following Feymann diagrams:

\vspace{0.25cm}

\epsfig{file=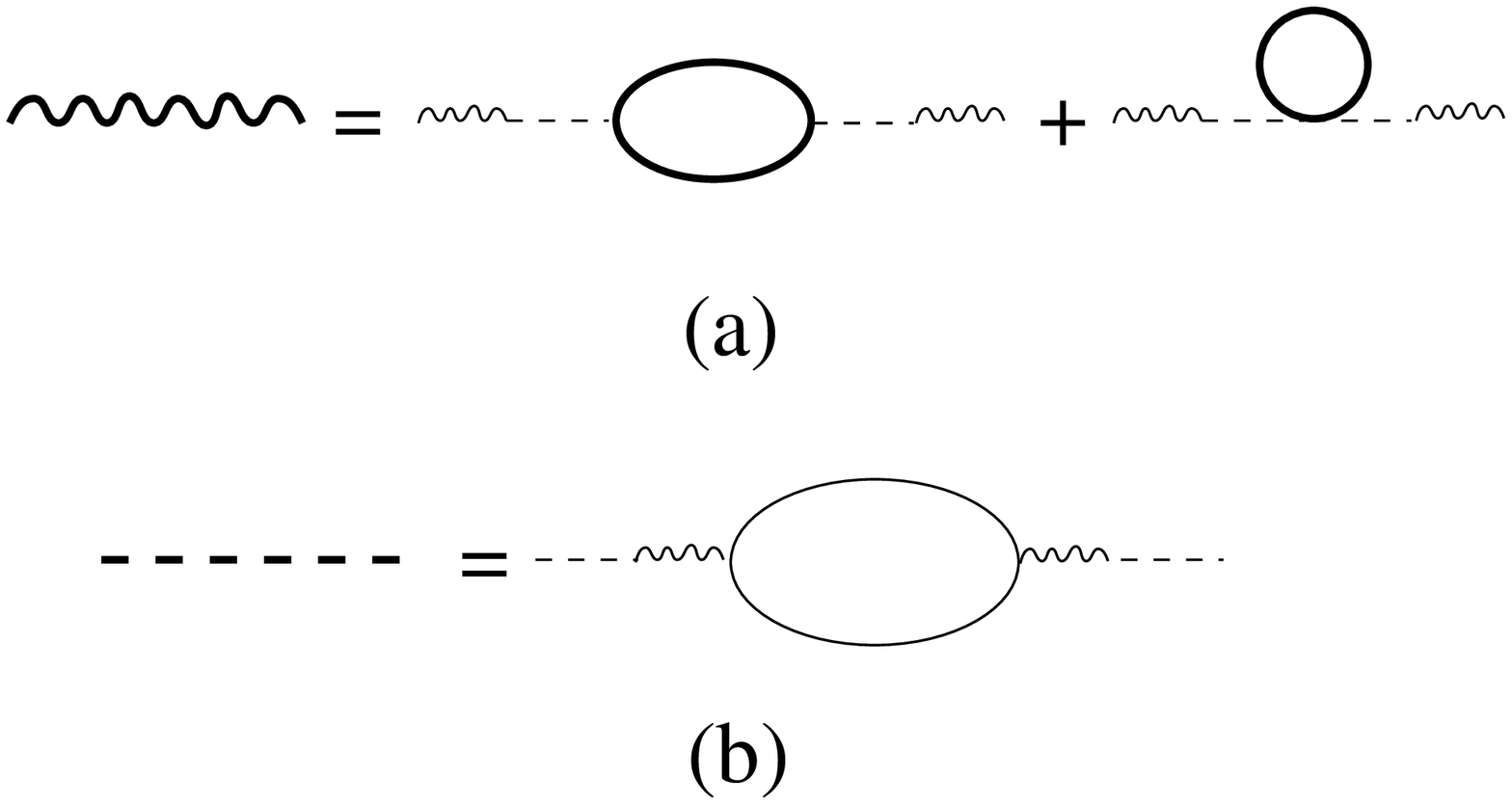,width=3.2in,height=0.9in,angle=0}

{\footnotesize {\bf Fig 2:} The renormalized propagators for $ a^{\psi} $ (thick wiggle line )
    and $ a^{\Phi} $ (thick dashed line). The thin wiggle line stands for $ a^{\psi} $,
   the dashed thin line for $ a^{\Phi} $, the thin solid line for fermion
    propagators, the thick solid line for the boson propagators.}

\vspace{0.25cm}

  By using the bare propagators $ <a^{\psi}_{\mu} a^{\Phi}_{\nu} > =-\epsilon_{\mu \nu \lambda}
     k_{\lambda}/k^{2} $ and the bare fermion and boson loop results
   $ \Pi^{f0}_{\mu \nu} = \Pi^{b0}_{\mu \nu} = -\frac{g^{2}}{16}
   k ( \delta_{\mu \nu} -k_{\mu} k_{\nu}/k^{2} ) $, we can find easily the
   renormalized $ a^{\psi} $ and $ a^{\Phi} $ propagators
    $ G^{\psi}_{\mu \nu} = G^{\Phi}_{\mu \nu} = -\frac{g^{2}}{16} \frac{1}{k}
    ( \delta_{\mu \nu} -k_{\mu} k_{\nu}/k^{2} ) $. In contrast to the conventional
   CS theory studied in \cite{chen,wen}, the propagators are {\em even} in $ k $,
   this is because the theory respects $ T $ symmetry. On the other hand, in contrast to
   the Maxwell propagators, they behave as $ 1/k $ instead of $ 1/k^{2} $. 

    The three loop diagrams for $ \Pi_{1} $ are given by:

\vspace{0.25cm}

\epsfig{file=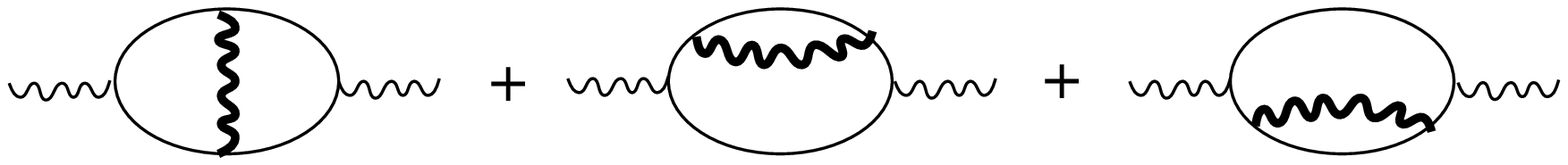,width=3.2in,height=0.6in,angle=0}

{\footnotesize {\bf Fig 3:} The three loop diagrams of $ \Pi_{1} $. The thick wiggle line
  stands for the renormalized propagators of $ a^{\psi} $,
  the thin solid line stands for the fermion propagator.  The one loop diagram is not shown.}    

\vspace{0.25cm}

   By using $ G^{\psi}_{\mu \nu} $ and
   extracting the symmetric part of the gauge propagator in the large
   $ N $ results in Refs.\cite{chen,wen}, we are able to calculate the above three
   loop diagrams.  Furry's theorem can be used to eliminate large number of null diagrams.
   We get the following series:
\begin{equation}
  \Pi_{1} = M N\frac{\pi}{8}
  (1+ \frac{3}{16} \frac{g^{4}}{ (2 \pi)^{2} } +  g^{8} + \cdots)
\end{equation}
  where  $ N=4 $ ( $ M=1 $ ) is due to the sum over 4 (1) species of Dirac 
  fermions ( bosons ).

    The three loop diagrams for $ \Pi_{3} $ are given by:

\vspace{0.25cm}

\epsfig{file=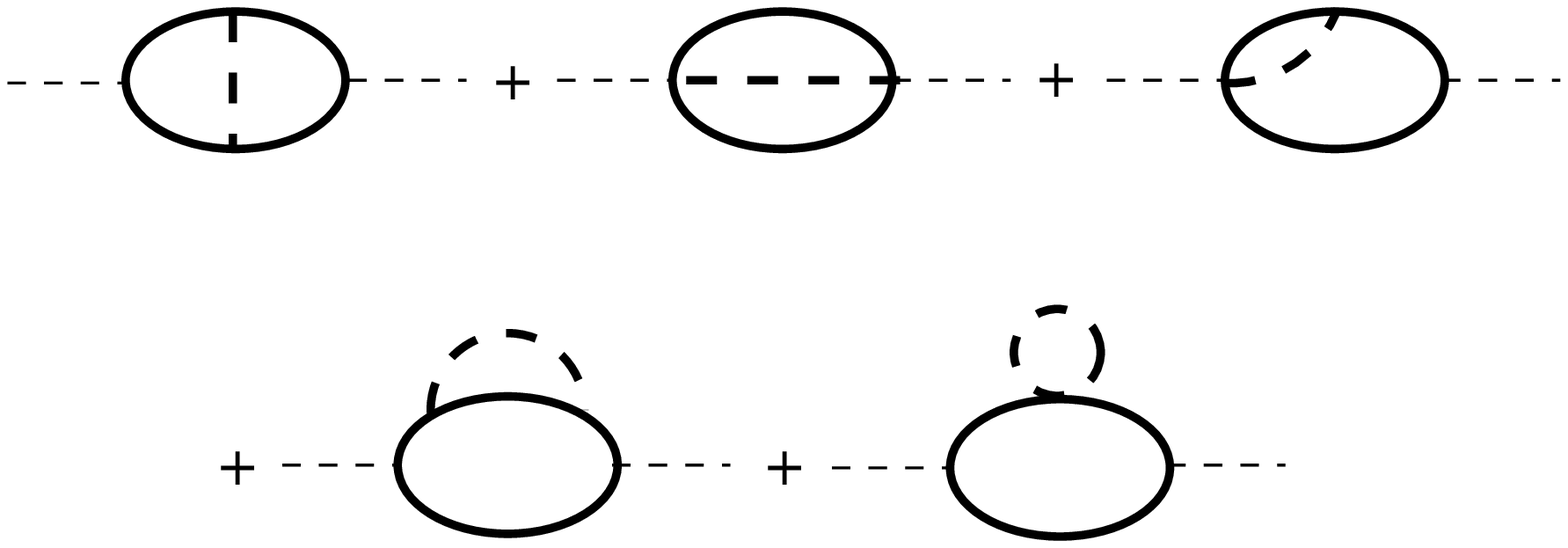,width=3.2in,height=1in,angle=0}

{\footnotesize {\bf Fig 4:} The three loop diagrams of $ \Pi_{3} $. The thick dashed line
  stands for the renormalized propagator of $ a^{\Phi} $,
  the thick solid line stands for the boson propagator.  The one loop diagram is not shown.}    

\vspace{0.25cm}

   For bosons, only one loop result is known:
\begin{equation}
   \Pi_{3} = M N \frac{\pi}{8} (1+  g^{4} +  g^{8} + \cdots)
\end{equation}
    Note that both $ \Pi_{1} $ and $ \Pi_{3} $ are 
   even functions of $ \theta $.
  
  From Furry's theorem, one of the first non-vanishing diagram
  for $ \Pi_{2} $ is:

\vspace{0.25cm}

\epsfig{file=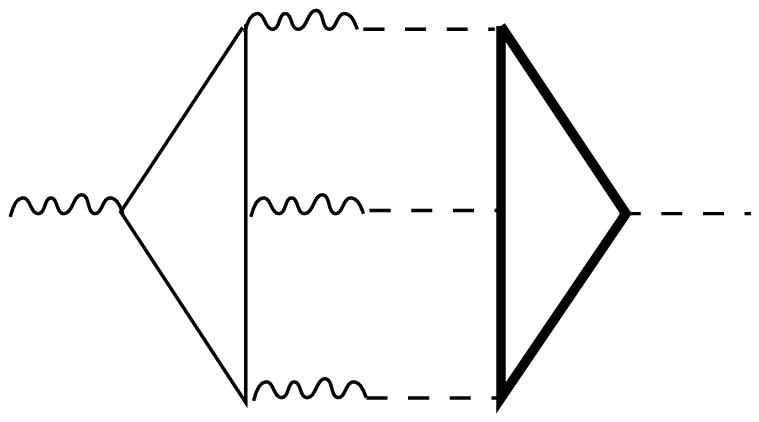,width=3.2in,height=0.8in,angle=0}

{\footnotesize {\bf Fig 5:} One of the four loop diagrams of $ \Pi_{2} $.
  the thin solid line stands for the fermion propagator,    
  the thick solid line for the boson propagator. All the other four loop diagrams can
  be obtained by shuffling the positions of the three bare propagator lines.}

\vspace{0.25cm}

  The series is $ \Pi_{2}= M N( g^{6} + g^{10} + \cdots) $ which
  is an {\em odd} function of $ \theta $.

  From Eqn.\ref{three}, $ \sigma^{\psi}, \sigma^{\Phi} $ are even functions
 of $ \theta $, but the mutual Hall drag conductivity is an {\em odd} function
 of $ \theta $. These are expected from $ P-H $ transformation.
 Under the $ P-H $ transformation of the
 vortex operator $ \Phi \rightarrow \Phi^{\dagger} $ in Eqn.\ref{mutual},
  it can be shown
  that $ A^{\Phi}_{\mu}, \theta $ is equivalent to $ -A^{\Phi}_{\mu}, -\theta $.
  From Eqn.\ref{matrix}, we reach the same conclusions. Specifically,
  $ \sigma^{H} $ takes opposite values for $ \theta=\pm 1/2 $,
  the periodicity under $ \theta \rightarrow \theta +1 $ is not preserved.
  Experimentally, the Hall drag conductivity can be detected by measuring
  the transverse voltage drop ( or transverse temperature drop for thermal
  conductivity ) of spinons due to the longitudinal driving of vortices.
  The Hall drag conductivity in double layer Quantum Hall systems has been
  investigated
  by several authors \cite{kun}. In double layer systems, the electrons
  in different layers are treated as two different species. There is a mutual
  CS interaction between the two species (both are fermions)
  which is directly responsible for
  this Hall drag conductivity, although the Coulomb interaction between the
  two species is responsible for the Coulomb ( longitudinal ) drag \cite{double}.
   This example shows that {\em no} external magnetic field is needed to produce
   a Hall effect ! For example $ \theta=\pm 1/2 $ lead to opposite Hall drag
   conductivities although static
  $ \alpha= \pm 1/2 $ vortex leads to {\em no} Hall conductivity.

   Again, by using the large $ N $ result of \cite{chen,wen},
  we find the anomalous dimensions of the fermion and vortex to two loops:

\vspace{0.25cm}

\epsfig{file=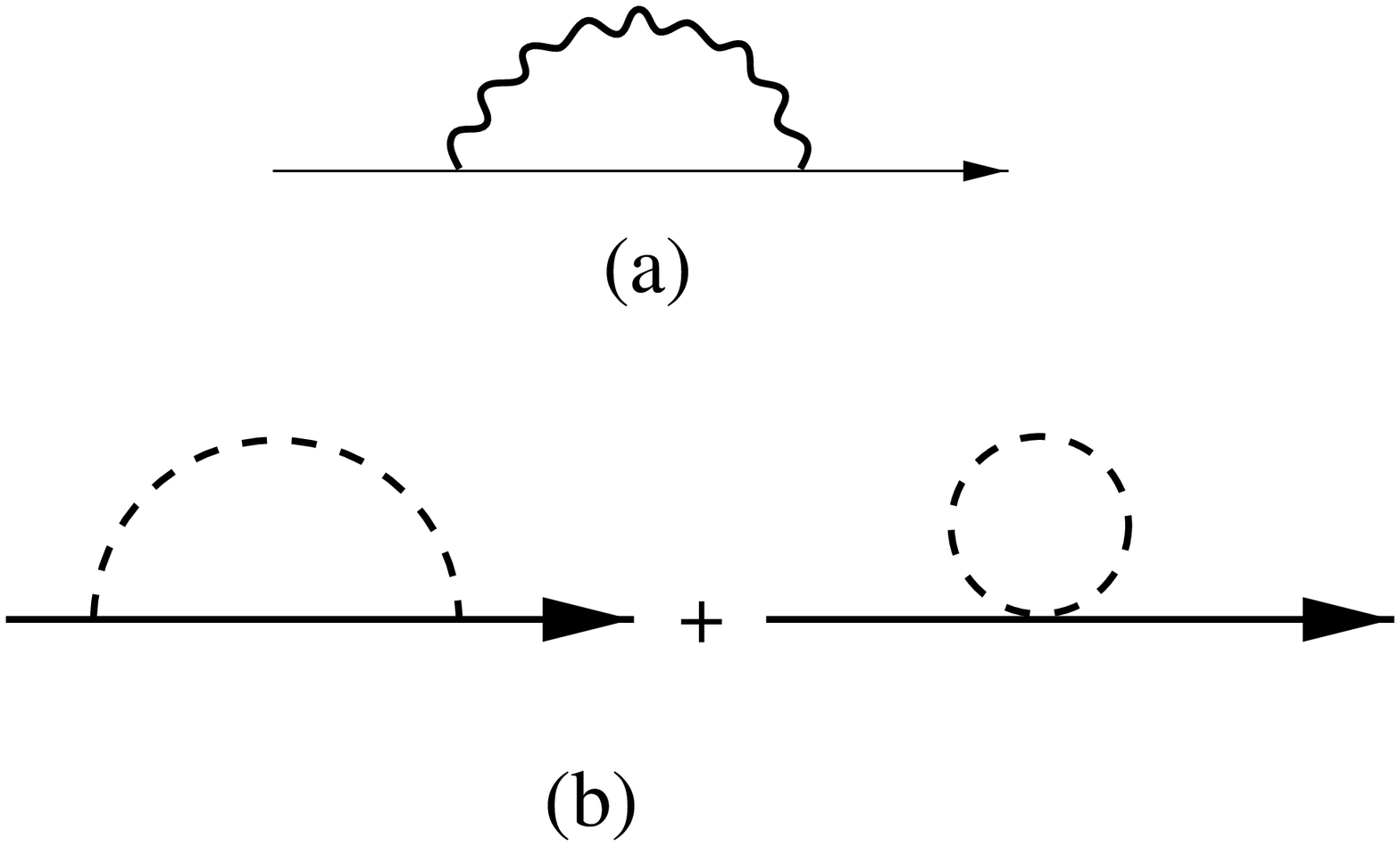,width=3.2in,height=0.9in,angle=0}

{\footnotesize {\bf Fig 6:} The two loop diagrams for the self energies of
  spinon (a) and vortex (b). The renormalized operators are used.}

\vspace{0.25cm}

     The results are:
\begin{eqnarray}
   \eta_{\psi} & = &  -\frac{g^{4}}{ 48 \pi^{2} } M  \nonumber \\
   \eta_{\Phi} & = & \eta_{XY}  -\frac{g^{4}}{ 12 \pi^{2} } N 
\end{eqnarray} 
  where $ \eta_{XY} \sim 0.038 $ is the anomalous
  dimension for the 3d XY model\cite{recent}. 

  The correlation length exponent can also be calculated to two loops by 
  the insertion of the operator $ \Phi^{\dagger} \Phi $:
\vspace{0.25cm}

\epsfig{file=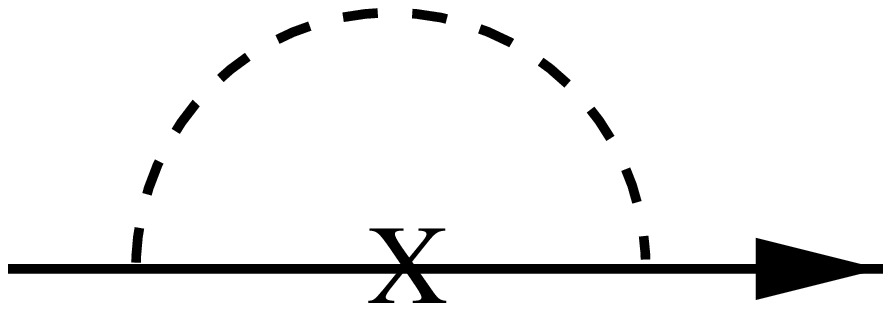,width=3.2in,height=0.8in,angle=0}

{\footnotesize {\bf Fig 7:} The cross stands for the operator insertion
  of $ \Phi^{\dagger} \Phi $.}

\vspace{0.25cm}

    The result is:
\begin{equation}
   \nu  =  \nu_{XY}  -\frac{g^{4}}{ 12 \pi^{2} } N
\end{equation}
  where $ \nu_{XY} \sim 0.672 $ is correlation length exponent for
  the 3d XY model\cite{recent}.

   It is instructive to go to dual representation of Eqn.\ref{mutual}, namely
  go to the boson representation: 
\begin{eqnarray}
   {\cal L} &= & \psi^{\dagger}_{a} \gamma_{\mu} (\partial_{\mu}
    -i a^{\psi}_{\mu} -i A^{\psi}_{\mu} ) \psi_{a}   
   +  | ( \partial_{\mu} -i a^{\phi}_{\mu} ) \phi |^{2}
      + V( |\phi|)  \nonumber  \\
    &  + &  \frac{1}{4} ( f^{\phi}_{\mu \nu})^{2}+
    i a^{\Phi}_{\mu} 
    \epsilon_{\mu \nu \lambda } \partial_{\nu} ( a^{\phi}_{\lambda}
          -a^{\psi}_{\lambda}/\theta)
    -i A^{\Phi}_{\mu} \epsilon_{\mu \nu \lambda} \partial_{\nu}
      a^{\phi}_{\lambda}
\label{mutuald}
\end{eqnarray}
  where $ V(|\phi|)= m^{2}_{\phi} |\phi|^{2}+ g_{\phi} |\phi|^{4} +\cdots $,

    Integrating out $ a^{\Phi}_{\mu} $ leads to the constraint up to a pure
  gauge:
\begin{equation}
          a^{\psi}_{\mu}= \theta a^{\phi}_{\mu}
\label{cons}
\end{equation}

      Substituting the above constraint to Eqn.\ref{mutuald} and setting
     $ a^{\phi}_{\mu} = a_{\mu} $, we find:
\begin{eqnarray}
   {\cal L} &= & \psi^{\dagger}_{a} \gamma_{\mu} (\partial_{\mu}
    -i \theta a_{\mu} -i A^{\psi}_{\mu} ) \psi_{a} 
     + \frac{1}{4}  f_{\mu \nu}^{2}     \nonumber  \\
  & + & | ( \partial_{\mu} -i a_{\mu} ) \phi |^{2} + V( |\phi|)
    -i A^{\Phi}_{\mu} \epsilon_{\mu \nu \lambda} \partial_{\nu}
      a_{\lambda}
\label{mutualb}
\end{eqnarray}

    The above Eqn. indicates that fermions and bosons are coupled to the
  same gauge field whose dynamics is described by Maxwell term instead of
  C-S term. This Eqn. is simply $ 2+1 $ dimensional combination of
  spinor QED and scalar QED. This action is similar to a relativistic analogue of
  the $ U(1) $ gauge field theory investigated extensively in Ref.\cite{u1}.
  The RG analysis at $ 4-\epsilon $ by dimensional
  regularization is possible, because the marginal dimension of all the relevant
  couplings are 4. However a RG analysis directly at $ 2+1 $ dimension is formidable.
   The physical meaning of $ \theta $ is obscure and the {\em exact marginality} of
  $ \theta $ is a highly non-trival result in the boson representation, but all these become
  evident in the dual vortex representation Eqn.\ref{mutual}.  It is evident that
  there is no periodicity under $ \theta \rightarrow \theta +1 $ in
  Eqn.\ref{mutualb}.

    If we perform duality  transformation again on Eqn.\ref{mutualb} to
  go to the vortex representation, then
  we recover Eqn.\ref{dual} upon neglecting the two Maxwell terms which
  are subleading to the mutual C-S term.

    In the next subsection, setting the two source terms vanishing, we
   look at the properties of the different phases on the two sides of
   this quantum critical point.

\subsection{ $ 2+ 1 $ dimensional QED in the underdoped phase  and 
             free fermions in the superconducting phase}

   In the disordered phase, the vortex condense $ < \Phi > = \Phi_{0}  $
   which generates a mass term for $ a^{\Phi}_{\mu} $ in Eqn.\ref{mutual}
\begin{equation}
  \frac{\Phi^{2}_{0}}{2} ( a^{\Phi}_{\mu} )^{2}_{t}
\end{equation}
     where the subscript $ t $ means transverse projection.

    Integrating out the massive $ a^{\Phi}_{\mu} $ leads to a Maxwell term
  for $ a^{\psi}_{\mu} $:
\begin{equation}
   {\cal L} =  \psi^{\dagger}_{a} \gamma_{\mu} (\partial_{\mu}
    -i  a^{\psi}_{\mu} ) \psi_{a} + \frac{1}{4 \Phi^{2}_{0}
} ( f^{\psi}_{\mu \nu} )^{2}
\label{QED}
\end{equation}

  This is simply $ 2+1 $ dimensional spinor QED which was studied
  by large N expansion by the authors in \cite{qed}. They  used four component
     Dirac fermions and discussed possible dynamic mass generation of
     the fermions which break the chiral symmetry, but preserves Parity and Time reversal symmetry.
     However, for $ 2 + 1 $ dimensional compact $ QED $, the possible important instanton effects were not
     well understood yet.
     The possible connection between this dynamic mass generation and Anti-ferromagnetism
     in the context of gapless flux phase was discussed by Marston \cite{brad}.

 In fact, we reach the same
  description from the boson representation Eqn.\ref{mutualb}. Because
  in the disordered phase, the boson $ \phi $ is massive, therefore can be
  integrated out, it generates the Maxwell term $ \frac{1}{4 m_{\phi}}
     f^{2}_{\mu \nu} $ which dominates over the existing {\em non-critical}
   Maxwell term. We reach Eqn.\ref{QED} after identifying $ m_{\phi} \sim \Phi^{2}_{0} $. 

     In the superconductor phase, the vortex $ \Phi $ is massive, therefore
  can be integrated out, it leads to the old Maxwell term 
    $  \frac{1}{4 m_{\Phi}} ( f^{\Phi}_{\mu \nu})^{2} $.
       Integrating out $ a^{\Phi} $ generates a mass term: 
\begin{equation}
   {\cal L} =  \psi^{\dagger}_{a} \gamma_{\mu} (\partial_{\mu}
    -i a^{\psi}_{\mu}) \psi_{a} 
     +\frac{m_{\Phi}}{2} (a^{\psi}_{\mu})_{t}^{2}
\label{free}
\end{equation}
    where $ m_{\Phi} $ is the mass of the vortex.

   The Dirac fermions become free.  In fact, we reach the
  same description from the boson representation Eqn.\ref{mutualb}. Because
   in superconductor phase, the boson $ \phi $ condense, therefore generates
   a mass term $ \frac{\phi^{2}_{0}}{2} (a_{\mu})^{2}_{t} $ which renders
   the Maxwell term ineffective. We reach the same conclusion from both
   sides by identifying $ m_{\Phi} \sim \phi^{2}_{0} $ which is dual to
   the relation in the disordered phase $ m_{\phi} \sim \Phi^{2}_{0} $.

    In short, in the disordered phase, the system is described by spinor
    QED Eqn.\ref{QED}; in the superconductor side, by free Dirac fermion. 
    We can view the transition as a simple example of
    confinement-deconfinement transition.
    In the confined ( disordered ) phase, the boson and fermion are confined 
    together by the fluctuating gauge field. In the deconfined ( superconductor)
    phase, the boson condensed, the gauge field becomes massive, the fermion becomes free. There is a line
    of fixed point governed by the mutual statistical angle $ \theta $
    separating the two phases.
\section{ The effect of charge fluctuation and the absence of massless dynamic gauge field at $ T=0 $ }

  In this section, we try to investigate the effect of charge fluctuation on
  the fixed line characterized by the statistical angle $ \theta $ discovered
  in the last section by considering the combined effects of
  $ U(1)_{cs} $ mutual statistical gauge fluctuation and $ U(1)_{c} $ electrical
  gauge fluctuation and treat both $ U(1) $ gauge fields on the equal footing.
  We find the condensation of
  $ hc/2e $ vortex condensation indeed leads to the confinement of
  spinon and chargon into Cooper pairs and electrons, in contrast to
  the condensation of $ hc/e $ vortex. The fixed line in the last section
   is destroyed and replaced by a quantum critical
   point separating the superconducting state and some sort of charge ordered
   state. There is {\em no} gapless gauge field left in the final theory,
   in sharp contrast to the $ U(1) $ or $ SU(2) $ gauge theory \cite{u1}.
   Both mutual C-S interaction term and the dynamic Doppler-shift term become
   irrelevant short-ranged interactions.
  
    The authors in Ref.\cite{guess} studied the similar model.
  They performed a perturbative RG calculation and reached very different
  conclusions from ours.

\subsection{ Quantum critical point}
  Putting $ a^{\Phi}_{\mu} \rightarrow 
      a^{\Phi}_{\mu} -a_{\mu} $ in Eqn.\ref{dual}, we get
\begin{eqnarray}
   {\cal L} &= & \psi^{\dagger}_{a} \gamma_{\mu} (\partial_{\mu}
    -i a^{\psi}_{\mu}) \psi_{a}  +
   | ( \partial_{\mu} -i a^{\Phi}_{\mu} ) \Phi |^{2}
   + V( |\Phi|)   \nonumber  \\
    & + & \frac{i}{ 2 \pi \theta} a^{\psi}_{\mu} 
    \epsilon_{\mu \nu \lambda } \partial_{\nu} a^{\Phi}_{\lambda}
      - i a_{\mu} \epsilon_{\mu \nu \lambda } \partial_{\nu} (
     a^{\psi}_{\lambda}/\theta + A^{eff}_{\lambda} ) +
     \frac{1}{4} f^{2}_{\mu \nu}
\end{eqnarray}

    Integrating out the electric gauge field $ a_{\mu} $ leads to:
\begin{eqnarray}
   {\cal L} &= & \psi^{\dagger}_{a} \gamma_{\mu} (\partial_{\mu}
    -i a^{\psi}_{\mu}) \psi_{a}  +
   | ( \partial_{\mu} -i a^{\Phi}_{\mu} ) \Phi |^{2}
   + V( |\Phi|)   \nonumber  \\
    & + & \frac{i}{ 2 \pi \theta} a^{\psi}_{\mu} 
    \epsilon_{\mu \nu \lambda } \partial_{\nu} a^{\Phi}_{\lambda}
     +\frac{1}{2} (a^{\psi}_{\mu}/\theta + A^{eff}_{\mu} )_{t}^{2}
\label{vortexm} 
\end{eqnarray}

    Comparing to Eqn.\ref{mutual},
   it is easy to see that the charge fluctuation leads to a mass term for
   the gauge field $ a^{\psi}_{\mu} $.
   Shifting $ a^{\psi}_{\mu}/\theta + A^{eff}_{\mu} \rightarrow a_{\mu} $ 
   and adding the gauge fixing term
    $ \frac{1}{2 \alpha} ( \partial_{\mu} a _{\mu} )^{2} $, we can integrate
    out the {\em massive } gauge field $ a_{\mu} $ in Lorenz gauge $
    \alpha=0 $ and find:
\begin{eqnarray}
   {\cal L} &= & \psi^{\dagger}_{a} \gamma_{\mu} (\partial_{\mu}
    -i \theta A^{eff}_{\mu}) \psi_{a}  +
   | ( \partial_{\mu} -i a^{\Phi}_{\mu} ) \Phi |^{2}
   + V( |\Phi|)   \nonumber  \\
    & + & \frac{1}{ 4} ( f^{\Phi}_{\mu \nu} )^{2} 
   -i ( A^{eff}_{\mu} + \theta j^{s}_{\mu} )
     \epsilon_{\mu \nu \lambda } \partial_{\nu} 
   a^{\Phi}_{\lambda} + (j^{s}_{\mu})^{2}_{t}
\end{eqnarray}

  Note the Maxwell term for $ a^{\Phi} $ is generated by the integration
  over the massive $ a_{\mu} $.

  Setting $ A_{\mu}=0 $, integrating out the fermions only leads to
 higher derivative terms than the Maxwell term:
\begin{equation}
   {\cal L} =  | ( \partial_{\mu} -i a^{\Phi}_{\mu} ) \Phi |^{2}
   + V( |\Phi|)   \nonumber  \\
     +  \frac{1}{ 4} ( f^{\Phi}_{\mu \nu} )^{2} + \cdots 
\end{equation}
  where $ \cdots $ means higher than second order derivatives.
 Therefore the vortex and fermion are asymptotically decoupled.
   It indicates that the charge fluctuation neglected in the last section
   destroy the fixed line characterized by $ \theta $. However,
   very different conclusions are reached in Ref.\cite{guess}.
   We think that the perturbative RG calculation in \cite{guess}
   may not treat the charge gauge field fluctuation correctly.

  Just like the last section, it is instructive to go to dual
  representation of Eqn.\ref{vortexm}, namely
  go to the boson representation: 
\begin{eqnarray}
   {\cal L} &= & \psi^{\dagger}_{a} \gamma_{\mu} (\partial_{\mu}
    -i a^{\psi}_{\mu} ) \psi_{a}   
   +  | ( \partial_{\mu} -i a^{\phi}_{\mu} ) \phi |^{2}
      + V( |\phi|)  \nonumber  \\
    &  + &  \frac{1}{4} ( f^{\phi}_{\mu \nu})^{2}+
    i a^{\Phi}_{\mu} 
    \epsilon_{\mu \nu \lambda } \partial_{\nu} ( a^{\phi}_{\lambda}
          -a^{\psi}_{\lambda}/\theta)  \nonumber  \\
     & +  &\frac{1}{2} (a^{\psi}_{\mu}/\theta + A^{eff}_{\mu} )_{t}^{2}
\label{charged}
\end{eqnarray}

   Integrating out $ a^{\Phi}_{\mu} $ leads to the {\em same}
  constraint as Eqn.\ref{cons} up to a pure gauge:
\begin{equation}
          a^{\psi}_{\mu}= \theta a^{\phi}_{\mu}
\end{equation}

      Substituting the above constraint to Eqn.\ref{charged} and setting
     $ a^{\phi}_{\mu} = a_{\mu} $, we find:
\begin{eqnarray}
   {\cal L} &= & \psi^{\dagger}_{a} \gamma_{\mu} (\partial_{\mu}
    -i \theta a_{\mu} ) \psi_{a}   
   +  | ( \partial_{\mu} -i a_{\mu} ) \phi |^{2}
      + V( |\phi|)  \nonumber  \\
   & + & \frac{1}{4}  f_{\mu \nu}^{2}
     +\frac{1}{2} (a_{\mu} + A^{eff}_{\mu} )_{t}^{2}
\label{chargeb}
\end{eqnarray}

   Comparing to Eqn.\ref{mutualb}, the important difference is that the gauge field
  acquires a mass due to the charge fluctuation which renders
  the Maxwell term ineffective. Note that the dynamic Doppler shift term is encoded in $ A^{eff}_{\mu} $.
  Up to irrelevant couplings, we can safely set $ a_{\mu}= A^{eff}_{\mu} $ in
  the above equation and find
\begin{equation}
   {\cal L} =  \psi^{\dagger}_{a} \gamma_{\mu} (\partial_{\mu}
    -i 2 \theta A_{\mu} ) \psi_{a}   
   +  | ( \partial_{\mu} -i 2 A_{\mu} ) \phi |^{2}
      + V( |\phi|) + \cdots 
\label{decouple}
\end{equation}
  where $ \cdots $ means the irrelevant couplings between bosons and fermions.

  The above Eqn. leads to the conclusion that
  the mutual statistical interaction between $ hc/2e $
  vortex and spinons leads to the confinement of spinon and chargon into
  electron and Cooper pair. The quasi-particles
  carry charge $2 \theta e $ which can be identified as
  electrons or holes, the Cooper-pairs are described by 3D X-Y model.
  The mutual C-S coupling and the dynamic Doppler-shift coupling between Cooper pairs and electrons
  are short-ranged interactions which are irrelevant near the quantum
  critical point. There is no gapless gauge fields in the final
  effective action ! However, there are several important factors left out
  in the above analysis: (1) The Berry phase term for the boson which
  is first order time derivative term is not included. This term is vanishing only at
  half-filling with particle-hole symmetry hopping. (2) The long-range 
  Coulomb interaction between Cooper pairs and electrons is not included.
  When the both effects are taken into account, the final ground state
  in the underdoped side at $ T=0 $ maybe  an insulating state with some
  kind of charge order. The nature of the QCP between the d-wave superconducting
  state and this charge ordered state is an important unsolved problem.
  However, the conclusion that there is no spin-charge separation and no
  gapless dynamic gauge fluctuation remains robust !

\subsection{ Disordered and superconducting phases}
  We follow the discussions in the previous section. In the disordered phase,
  the vortex condense $ < \Phi > = \Phi_{0}  $
  which generates a mass term for $ a^{\Phi}_{\mu} $ in Eqn.\ref{vortexm}.
  Integrating out the massive $ a^{\Phi}_{\mu} $ leads to a Maxwell term
  for $ a^{\psi}_{\mu} $.
\begin{equation}
   {\cal L} =  \psi^{\dagger}_{a} \gamma_{\mu} (\partial_{\mu}
    -i a^{\psi}_{\mu}) \psi_{a} 
     +\frac{1}{4 \Phi^{2}_{0}} (f^{\psi}_{\mu \nu})^{2}
     +\frac{1}{2} (a^{\psi}_{\mu}/\theta + A^{eff}_{\mu} )_{t}^{2}
\label{massive}
\end{equation}
   Because of the mass term, the Dirac fermions become free and carry spin $ 1/2 $ and charge
   $ 2 \theta e $. In fact, we reach the same conclusion from the boson representation
   Eqn.\ref{chargeb}.

   In the superconductor phase, the vortex $ \Phi $ is massive, therefore
   can be integrated out, it leads to the old Maxwell term 
   $  \frac{1}{4 m_{\Phi}} ( f^{\Phi}_{\mu \nu})^{2} $.
   Integrating out $ a^{\Phi} $ generates a mass term for
   $ a^{\psi}_{\mu} $:
\begin{equation}
   {\cal L} =  \psi^{\dagger}_{a} \gamma_{\mu} (\partial_{\mu}
    -i a^{\psi}_{\mu}) \psi_{a} 
     +\frac{m_{\Phi}}{2} (a^{\psi}_{\mu})_{t}^{2}
     +\frac{1}{2} (a^{\psi}_{\mu}/\theta + A^{eff}_{\mu} )_{t}^{2}
\label{con}
\end{equation}
    where $ m_{\Phi} $ is the mass of the vortex.

   Diagonizing the last two mass terms leads to a {\em continuously changing
  charge }. This is expected, because in the superconducting state,
  the quasi-particle is a linear 
  combination of electrons and holes, therefore carrying continuously
  changing charge.
  In fact, we reach the same description from the boson representation
   Eqn.\ref{chargeb}. Because
   in superconductor phase, the boson condense $ <\phi> =\phi_{0} $,
   therefore generates a mass term for $ a_{\mu} $:
\begin{equation}
   {\cal L} =  \psi^{\dagger}_{a} \gamma_{\mu} (\partial_{\mu}
    -i \theta a_{\mu} ) \psi_{a}   
    +  \frac{\phi^{2}_{0}}{2} (a_{\mu} )_{t}^{2}
     +\frac{1}{2} (a_{\mu} + A^{eff}_{\mu} )_{t}^{2}
\end{equation}
     Which is essentially the same as Eqn.\ref{con} after the identification
   $ m_{\Phi} \sim \phi^{2}_{0} $.

     In short, in the disordered phase, the quasi-particles carry charge
  $ 2 \theta e $ and spin  $ 1/2 $ which can be identified as
  electrons or holes. When the Berry phase term and the long-range
  Coulomb interactions are taken into account, the electrons may form
  some charge ordered state.
  In the superconducting side, the quasi-particles
  carry continuously changing charge which is a linear combination
  of electrons and holes.

\section{ Application to static disordered vortex array and thermally generated vortices}

  So far, we discussed the vortices generated by quantum fluctuations in the underdoped regime.
  Vortices can also be generated by an external magnetic field inside superconductors or
  by thermal fluctuations in the pseudo-gap regime. They were discussed in Refs.\cite{static} and \cite{tough}
  respectively. In this section, we apply the formalism developed for quantum generated vortices
   at $ T=0 $ to study vortices generated by an external magnetic field or by thermal fluctuations. By studying
  the three different kinds of vortices in a unified picture, we not only
  recover the previous established results, but also provide additional physical insights
  on the internal gauge field fluctuations.

  Inside the $d$-wave superconductor, although the Cooper-pair condensates dominates the electrical
  transport at zero frequency, the quasi-particles near the four nodes of a $d$-wave superconductor
  are responsible for the heat and spin transport at low temperature. Recently Ye studied the quasi-particle
  transport in a random vortex array with a fully quantum mechanical approach  \cite{static}.
  Although the previous semi-classical approach only capture the physics of the Volovik effect,
  fully quantum mechanical approach treats the Volovik effect ( scalar field scattering ) and
  the AB phase scattering ( the gauge field scattering ) on the equal footing.
  Ye found that the long range logarithmic
  interaction between vortices suppress the superfluid velocity fluctuation,
  but does not affect the internal gauge field fluctuation. He concluded
  that the quasi-particle scattering from the random gauge field dominate
  over that from the superfluid velocity ( the Volovik effect).

   Roughly speaking, the mixed state with vortex density $ n_{v} $ corresponds to the vortex
  condensed phase(disordered phase).
  As discussed in the previous sections, the vortex condensation leads to a mass term
  $ \frac{1}{2} n_{v} ( a^{\Phi}_{\mu} )^{2} $
  for the statistical gauge field $ a^{\Phi}_{\mu} $ in Eqn.\ref{dual}.
  Adding the mass term to the equation, integrating out $ a^{\Phi}_{\mu} $ and keeping only the time component of
  the charge gauge field $ a_{0} $ which mediates the long-range {\em density-density}
  logarithmic interaction between the static vortices,
  we get two {\em static} Maxwell terms for $ a^{\psi}_{\alpha}, v_{\alpha} $ respectively
  and a mass term for $ v_{\alpha} $:
\begin{eqnarray}
   {\cal L} & = &  \psi^{\dagger}_{a} \gamma_{\mu} (\partial_{\mu}
    -i a^{\psi}_{\alpha} ) \psi_{a}  + \sum_{\alpha} v_{\alpha} \psi^{\dagger}_{a}\psi_{a}   \nonumber  \\
   & + &  \frac{p}{ 4 n_{v} } ( f^{\psi}_{ \alpha \beta})^{2} 
     +  \frac{p}{ 4 n_{v} } ( f^{v}_{ \alpha \beta})^{2} + \frac{p}{ 4 } (v_{\alpha})^{2}  
\label{static}
\end{eqnarray}
  where $ p $ is added to stand for an unknown function of pinning potential,
  $ f^{\psi}_{\alpha \beta}= \partial_{\alpha} a^{\psi}_{\beta}-\partial_{\beta} a^{\psi}_{\alpha},
   f^{v}_{\alpha \beta}= \partial_{\alpha} v_{\beta}-\partial_{\beta} v_{\alpha} $.

   In Eqn.\ref{static}, the averages over $ a^{\psi}_{\alpha}, v_{\alpha}$
  should be understood as quenched instead of annealed averages. Namely there is no feedback effects on the gauge field and
  scalar field propagators from the fermions \cite{feed}.
  The above equation shows that the long-range logarithmic interaction
  between vortices suppress the fluctuation of the superfluid velocity and leads to a mass term for the superfluid velocity
  and the {\em incompressibility} of the vortex system. The quasi-particle is moving in a long-range
  correlated random magnetic field and short-range correlated scaler potential.
  The distribution of the random gauge field is given by the corresponding static Maxwell term 
  and that of the scalar field  is given by the corresponding static Maxwell term and the mass term.
  Eqn. \ref{static} is essentially the same as Eqn. 9 in Ref. \cite{static}.

  Strictly speaking, the gauge field from the random vortex array is a {\em random} $ Z_{2} $ gauge field.
  In Ref.\cite{static}, in order to perform an analytic calculation in the continuum limit,
  Ye made an approximation by replacing the discrete  random $ Z_{2} $ AB phase by a continuous
  random $ U(1) $ AB phase and performed a standard diagrammatic perturbation theory.

  In Ref.\cite{tough}, assuming the finite temperature phase transition 
  from $ d$-wave superconductor to pseudo-gap regime is a Kosterlize and Thouless (KT) vortex and anti-vortex
  unbinding transition \cite{emery}, Ye studied the electron spectral function
  in the vortex plasma regime in Fig.1 \cite{tough}. Starting from Eqn.\ref{tough}, he found  that the random gauge field
  $ a^{\psi } $ missed in the previous semiclassical approaches \cite{andy} destroyed the coherent spinon motion
  and leads to branch cut singularities and pertinent non-Fermi liquid behaviors.
  There are three subtle differences between thermally generated vortices from
  the static vortices generated by an external magnetic field
  inside the $d$-wave superconductor:
  (1) The vortex density $ n_{v} $ should be replaced by free vortex density $ n_{f} $ above KT transition temperature.
  (2) Thermally generated vortices are moving around, the  $ U(1) $ character is stronger than $ Z_{2} $ character. They destroy the superconductivity. But static vortices only destroy the superconductivity
  inside the vortex cores.
  (3) The distribution of thermally generated vortices is precisely given by vortex hydrodynamics in the long-wavelength limit.
      The free vortex density $ n_{f} $ is the only relevant parameter.  We can set $ p $ equal to 1 in Eqn.\ref{static}.

  After taking into account these subtle differences, the above
  procedures leading to Eqn.\ref{static} can be equally applied to thermally generated free vortices with density $ n_{f} $.
  The following alternative derivation of Eqn.\ref{static} may be stimulating:
  the vortices being treated classically, their commutation relations can be neglected, the dual singular gauge
  transformation in Eqn.\ref{mcs2} on the vortices is not necessary, the Berry phase term for
  the boson ( the linear time derivative term) can also be neglected.
  More specifically, only the time component of the {\em charge} gauge field
  $ a_{0} $ in Eqn.\ref{dual} is kept to mediate the long-range {\em density-density}
  logarithmic interaction between the vortices. Obviously, $ a_{0} $ couples
  to the vortices the same way as the superfluid velocity $ v_{\alpha} $
  couples to the spinon,
  its fluctuation leads to a mass term for the $ v-v $ correlation.
  However, being orthogonal to the spatial component
  $ \vec{a}^{\psi} $, $ a^{\psi}-a^{\psi} $
  correlator remains gapless. This also leads to exactly the same equation as Eqn.\ref{static} with $ p $ setting equal to
  1 and the same conclusions as those first reached in Ref.\cite{static}.

  However, at zero temperature,
  the vortices being treated quantum mechanically, all the components
  of the {\em charge} gauge field $ a_{\mu} $ should be kept to 
  mediate the long-range
  {\em current-current} logarithmic interactions between the vortices.
  Then the charge gauge field fluctuation leads to mass terms not only
  for $ v $, but also for the spatial component $ \vec{a}^{\psi} $.
  There is {\em no gapless dynamic } gauge fluctuations in the Cooper-pair
  picture as demonstrated in Eqn.\ref{massive}.
  The phenomenon that the two gapless gauge fields cancel each other is similar to the
  standard Higgs mechanism. In Higgs mechanism, the gapless gauge field is eaten by
  gapless Goldstone modes. The two gapless modes annihilate each other, the Goldstone mode disappears, while
  the gapless gauge field becomes massive.    

\section{discussions and conclusions}

 Based on the earlier work by Balents, Fisher and Nayak \cite{balents},
 Senthil and Fisher developed $ Z_{2} $
 gauge theory \cite{z2,guess} to study quasi-particles
 coupled to vortices generated by quantum fluctuations.
 By breaking electrons and Cooper pairs into smaller constitutes: chargons with
 charge $ e $, spin $ 0 $ and spinons with charge $ 0 $,spin $ 1/2 $,
 SF introduced a local $ Z_{2} $ gauge degree of
 freedom to constrain the
 Hilbert space to be the original one. The effective
 action describes both chargons and spinons coupled to local fluctuating
 $ Z_{2} $ gauge theory with a doping dependent Berry phase term.
 By the combination of standard duality transformation of 3 dimensional
 XY model and that of $ Z_{2} $ gauge theory, the action is mapped into a
 dual vortex representation where the $ hc/2e $ vortices and spinons 
 are coupled by a mutual $ Z_{2} $ CS gauge theory. As usual
 vortices in XY model, the $ hc/2e $ vortices also couple to a fluctuating
 $ U(1) $ gauge field which mediates the long-range logarithmic interaction
 between the vortices. Starting from the dual representation,
 the authors in Ref.\cite{guess}
 studied a transition from d-wave superconductor to confined Mott
 insulator driven by
 the condensing of $ hc/2e $ vortices at half filling. In order to study the critical
 behaviors of this particular confinement and de-confinement transition,
 they replaced the $ Z_{2} $ mutual CS theory on the lattice by $ U(1) $
 mutual CS theory in the continuum and performed renormalization Group (RG) 
 analysis. Some of their RG analysis may not be correct as
 demonstrated in Sec.IV.
  In contrast to $ U(1) $ gauge theory which has only confined phase, $ Z_{2} $ gauge theory has
  both confined  and deconfined phases. The topological excitation of the $ Z_{2} $ gauge field
  play an important role. If $ hc/2e $ vortices condense, the visons are also condensed,
  the system is in the confined phase, the chargon
  and spinon are confined into electron, Cooper-pair and magnon. If $ hc/e $ vortices condense,
  the visons are gapped, the system is in the deconfined phase, the spinon and chargon are asymptotically
  decoupled in the long-wavelength limit.  SF argued that the cuprates are in the deconfined phase,
  therefore the zero temperature transition at $ x=x_{c} $ is driven by the condensation of $ hc/e $ vortex.
  They further proposed the " vison trapping experiment " to test their spin-charge separation scenario.
  Several groups performed such experiments, but found no signature of visons \cite{kirtley}.

  The original Anderson singular gauge transformation \cite{and} was proposed
  for static vortices. In this paper, we extend the one singular gauge
  transformation for static vortices to two mutual flux-attaching singular gauge transformations
  for moving vortices generated by quantum fluctuations. By making a close
  analogy to the conventional singular gauge transformation of FQH system, we
  perform the two singular gauge transformations
  attaching flux of moving vortices to quasi-particles or vice versa. 
  Just like conventional singular gauge transformation leads to
  conventional CS term, the two mutual singular gauge transformations
  lead to mutual CS term. In this way, we propose an intuitive and
  physical transparent approach to bring out explicitly
  the underlying physics associated with the condensation of the
  $ hc/2e $ vortices.  When considering both mutual C-S interaction
  and long-range logarithmic interaction between the vortices, we find that
  there are only electrons and Cooper pairs in the spectrum and
 there are no gapless gauge field fluctuation in contrast
 to the $ U(1) $ or $ SU(2) $ gauge theory \cite{u1}.
 When taking into account both the linear time derivative Berry phase term for the bosons
 and the long-range Coulomb interactions between Cooper pairs and electrons, the true ground state
 in the underdoped side may be some unknown charge ordered state.
 Understanding the zero temperature transition from d-wave superconductor to
 this unknown charge ordered state is an important future research direction. 
  The electron spectral weight starting from Eqn.\ref{tough}
  measured by ARPES in the pseudo-gap regime at $ T> T_{c} $ indicated in Fig.1
  was presented in a separate publication \cite{tough}.
 The theory presented in this paper focus on the confined phase.
  The recent experiment \cite{kirtley} indicates that the cuprates may be inside the confined phase.
  In principle, our theory should be equivalent to the $ Z_{2} $ gauge theory in the confined phase side
  after tracing out
  the fluctuating $ Z_{2} $ gauge field except the Doppler shift effect missing in $ Z_{2} $ gauge field
  is also explicitly incorporated into the present approach.

  The importance of the Volovik effect and the AB phase scattering in random static vortex array was stressed
  in Ref.\cite{static}. As concluded in this paper, when the vortices are generated by quantum fluctuations at $ T=0 $,
  the dynamic Doppler shift effect and the mutual C-S phase scattering both become irrelevant short-ranged interactions.
  However, there is a crossover regime between these two limits which is the vortex plasma regime near the
  {\em finite temperature} KT transition. In this regime, there is a energy scale separation between the quasi-particle
  and the vortices: the vortices are moving much slower than the quasi-particles, therefore can be treated
  in a hydrodynamic approximation. The important effects of Volovik and AB phase scattering on ARPES data was discussed
  in Ref.\cite{tough}. 

  As explicitly stressed in this paper, the $ U(1) $
  mutual CS theory in the continuum does not have the periodicity under
  $ \theta \rightarrow \theta +1 $, because moving $ hc/2e $ vortex is different
  from moving $ -hc/2e $ vortex. 
  In fact, the periodicity of the conventional CS theory under
  $ \theta \rightarrow \theta+2 $ is also a very intricate issue.
  On the one hand,
  the perturbative RG expansion in terms of the statistical angle $ \theta $ in
  the continuum limit in Refs.\cite{wen,boson,subir} does not
  have this periodicity.
  On the other hand, properly regularized on the lattice,
  the CS theory does have
  this periodicity \cite{kiv}. 
  The two different regularizations do lead to different conclusions on the
  Quantum Hall transitions. It also leads to notorious Hall conductivity
  difficulty at $ \nu=1/2 $ \cite{not}. A deep understanding on the periodicity problem
  in both conventional CS term and mutual CS term is clearly required in the near future.
  
 This work was supported by NSF Grant No. PHY99-07949 at
 Institute for Theoretical Physics, University of California,
 Santa Barbara, CA, 93106 and the Pennsylvania state University.
  I am deeply indebted to A. Millis for numerous discussions. I am very grateful
  for J. K. Jain for discussions on the periodicity of C-S term.
 I also thank M. P. A. Fisher, S. Girvin, S. Kivelson, C. Lannert,
 P. A. Lee, S, Sachdev, T. Senthil and
 Kun Yang for helpful discussions.

\appendix

\section{ The most general form of the gauge propagators}

    Adding the gauge fixing terms $ \frac{1}{2 \alpha}
   ( (\partial_{\mu} a^{\psi}_{\mu} )^{2}+
    (\partial_{\mu} a^{\Phi}_{\mu} )^{2}) $ to Eqn.\ref{mutual}, we
   can find the gauge field propagators by inverting the matrix 
\begin{equation}
             \left( \begin{array}{cc}
    \Pi_{1}(k) k ( \delta_{\mu \nu} - \frac{ k_{\mu} k_{\nu} }{k^{2}} )
     + \frac{ k_{\mu} k_{\nu} }{\alpha} &
    \Pi_{2}(k) \epsilon_{\mu \nu \lambda} k_{\lambda}   \\
    \Pi_{2}(k) \epsilon_{\mu \nu \lambda} k_{\lambda}   &
    \Pi_{3}(k) k ( \delta_{\mu \nu} - \frac{ k_{\mu} k_{\nu} }{k^{2}} )
     + \frac{ k_{\mu} k_{\nu} }{\alpha} \\
			\end{array}   \right )
\end{equation}
    
     The results are:
\begin{eqnarray}
    (G_{\psi \psi})_{\mu \nu} & = & \frac{ \alpha k_{\mu} k_{\nu} }{ k^{4} }
    + \frac{\Pi_{3}}{ \Pi_{1} \Pi_{3} + \Pi^{2}_{2} } \frac{1}{k}
    ( \delta_{\mu \nu} - \frac{ k_{\mu} k_{\nu} }{k^{2}} ) 
               \nonumber  \\
    (G_{\psi \Phi})_{\mu \nu} & = & (G_{\Phi \psi})_{\mu \nu} = 
     \frac{\Pi_{2}}{ \Pi_{1} \Pi_{3} + \Pi^{2}_{2} } (- \frac{
     \epsilon_{\mu \nu \lambda} k_{\lambda} }{k^{2}} )
                    \nonumber  \\
    (G_{\Phi \Phi})_{\mu \nu} & = & \frac{ \alpha k_{\mu} k_{\nu} }{ k^{4} }
    + \frac{\Pi_{1}}{ \Pi_{1} \Pi_{3} + \Pi^{2}_{2} } \frac{1}{k}
    ( \delta_{\mu \nu} - \frac{ k_{\mu} k_{\nu} }{k^{2}} )  
 \end{eqnarray}

    In the Landau gauge $ \alpha=0 $ and putting $ \vec{k}=0 $, we recover
   the results calculated in Sec. III.

\section{ Moving $ hc/e $ vortices and the irrelevance of the dynamic Doppler shift effect }

  For $ hc/e $ vortex, we can perform a
  neutral-like single-valued gauge transformation by
  setting $ \phi_{A}= \phi_{B}= \phi/2 $.
 The double strength vortex and its stability was investigated in
 Ref.\cite{own}, the associated spin-charge separation
  has been discussed extensively in Refs.\cite{balents,z2}. In this appendix,
  considering a moving $ hc/e $ may still scatter quasi-particles by the AB phase,
  in analogy to the $ hc/2e $ vortices discussed in the main text,
  we still include a possible mutual C-S interaction with the statistical angle
   $ \theta $ taking {\em any } integer values.
  For $ \theta =0 $, we recover the results in Ref.\cite{balents}
  and also get the new result on the expression of current operator $ I_{\mu} $ in the
  superconducting phase. Although the static Volovik effect
  is marginally relevant \cite{static,tough} and cause finite density of states of the quasi-particles at zero energy.
  In this appendix, we show that surprisingly the dynamic Volovik effect is highly irrelevant on the both sides of
  the QCP. 

     Putting $ \phi_{A}=\phi_{B}=\phi/2 $, namely setting $ a_{\mu}=0 $
    in Eqn.\ref{act} leads to: 
\begin{eqnarray}
   {\cal L} &= & \psi^{\dagger}_{1a} [ \partial_{\tau} 
    + v_{f} p_{x} \tau^{3} + v_{\Delta} p_{y} \tau^{1} ]
    \psi_{1a} +  (1 \rightarrow 2, x \rightarrow y )
    \nonumber   \\
     & + & \frac{K}{2} ( \partial_{\mu} \phi - A^{eff}_{\mu} )^{2}
\label{acta}
\end{eqnarray}
    where the effective gauge field is $ A^{eff}_{\mu}= A_{\mu}-
    K^{-1}J_{\mu} $. There is no internal gauge
    field $ a_{\mu} $,
   because {\em static} double strength vortices do not scatter the quasi-particles by AB phase.

    Performing duality transformation to vortex representation leads to
\begin{eqnarray}
   {\cal L} &= & \psi^{\dagger}_{a} \gamma_{\mu} \partial_{\mu} \psi_{a}  +
   | ( \partial_{\mu} -i a_{\mu} ) \Phi |^{2}
   + V( |\Phi|)   \nonumber  \\
    & + & \frac{1}{ 4}  f_{\mu \nu}^{2}
   -i  A^{eff}_{\mu} \epsilon_{\mu \nu \lambda } \partial_{\nu}
   a_{\lambda} -\mu \epsilon_{ij} \partial_{i} a_{j}
\label{double}
\end{eqnarray}
   where the last ( Berry phase ) term can be absorbed into  $ A^{eff}_{\mu} $
  by redefining $ A^{eff}_{\mu} \rightarrow A^{eff}_{\mu} +
   i \mu \delta_{\mu 0} $.
 Comparing with Eqn.\ref{dual}, there is only charge fluctuation.
 There is no mutual C-S term, namely $\theta=0 $.

    The duality transformation to boson representation is
\begin{eqnarray}
   {\cal L} &= & \psi^{\dagger}_{a} \gamma_{\mu} \partial_{\mu} \psi_{a}
   +  | ( \partial_{\mu} -i a^{\phi}_{\mu} ) \phi |^{2}
      + V( |\phi|)  \nonumber  \\
    &  + &  \frac{1}{4} ( f^{\phi}_{\mu \nu})^{2}+
    i a_{\mu}
    \epsilon_{\mu \nu \lambda } \partial_{\nu} ( a^{\phi}_{\lambda}
          -A^{eff}_{\lambda} )
    + \frac{1}{4} f_{\mu \nu}^{2}
\end{eqnarray}

    Integrating out the electric gauge field leads to
\begin{eqnarray}
   {\cal L} &= & \psi^{\dagger}_{a} \gamma_{\mu} \partial_{\mu} \psi_{a}
   +  | ( \partial_{\mu} -i a^{\phi}_{\mu} ) \phi |^{2}
      + V( |\phi|)      \nonumber  \\
    &  + &  \frac{1}{4} ( f^{\phi}_{\mu \nu})^{2}
    + \frac{1}{2} ( a^{\phi}_{\mu} -A^{eff}_{\mu})^{2}_{t}
\end{eqnarray}

    Setting  $ a^{\phi}_{\mu}= A^{eff}_{\mu} $ up to irrelevant term leads to
\begin{equation}
   {\cal L}  =  \psi^{\dagger}_{a} \gamma_{\mu}   \partial_{\mu}  \psi_{a}
    +  \mu \phi^{\dagger} \partial_{\tau} \phi
   +  | ( \partial_{\mu} -i  A_{\mu} ) \phi |^{2}
      + V( |\phi|) + \cdots 
\label{old}
\end{equation}
  where the second term is the  Berry phase term which is a first order
  time derivative and $ \cdots $ means the irrelevant couplings between bosons and fermions.
  Eqn.\ref{old} is essentially the same as Eqn.\ref{acta}.

    However, a {\em moving} $ hc/e $ vortex may still scatter quasi-particles by AB phase.
    In general, we should also include a possible mutual C-S term in the effective action.
    The most general form is again given by Eqn.\ref{dual} with the statistical
    angle $ \theta $ taking any integers and $ A^{eff}_{\mu}= A_{\mu}- K^{-1}J_{\mu} $
    for $ hc/e $ vortex. The formulation developed in Sec.IV can be straight-forwardly
    applied to $ hc/e $ vortices.
    Putting  the above two values for $ \theta $ and $ A^{eff}_{\mu} $ into
    Eqn.\ref{chargeb} leads to
\begin{eqnarray}
   {\cal L} & = & \psi^{\dagger}_{a} \gamma_{\mu}  ( \partial_{\mu} -i \theta A_{\mu} ) \psi_{a}
    + \theta \mu \psi^{\dagger}_{a} \gamma_{0}   \partial_{0} \psi_{a}
                           \nonumber   \\
   & + & \mu \phi^{\dagger} \partial_{\tau} \phi
   +  | ( \partial_{\mu} -i  A_{\mu} ) \phi |^{2}
      + V( |\phi|) + \cdots 
\end{eqnarray}
  where the second and third terms are the Berry phase terms which are first order
  time derivatives.

   The above equation indicates that the quasi-particle carries charge $ \theta  $
   which can take any integer numbers. Obviously, there is {\em no}
   periodicity in $ \theta $. The most natural and simplest choice
   is $ \theta =0 $ where the above equation reduces to Eqn.\ref{old}. In this equation,
   bosons and fermions are asymptotically decoupled. The
   fermions (spinons) carry  only spin 1/2, the bosons ( holons)
   carry only charge $ e $.  There is spin-charge separation.
   The long-range Coulomb interaction could be incorporated by adding the dynamic term
   for the time component of the gauge field $ \frac{1}{2} k | A_{0}(k) |^{2} $ \cite{coul}.
   Without the fermionic part,
   the action is the same as the superconductor to insulator transitions
   studied in \cite{sit}, except here condensed is a charge $ e $ boson instead of a Cooper pair.

   In the underdoped regime, the vortex condenses $ < \Phi > = \Phi_{0}  $
   which generates a mass term for $ a_{\mu} $ in Eqn.\ref{double}
   $ \frac{1}{2} \Phi^{2}_{0} ( a_{\mu} )^{2}_{t} $.
    Integrating out the massive $ a_{\mu} $ leads to a Maxwell term
  for $ A^{eff}_{\mu} =A_{\mu}-K^{-1} J_{\mu} $ from which we can identify
  the automatically conserved electric current 
   $ I_{\mu} = \frac{1}{ \Phi^{2}_{0}}
   ( \partial^{2}_{\nu} J_{\mu}
   - \partial_{\mu} \partial_{\nu} J_{\nu} ) $ \cite{balents}. The {\em dynamic} Doppler shift effect
  is $ \sim \frac{1}{\Phi^{2}_{0}} (\partial J)^{2} $, it is a four fermion term with two
  derivative, therefore highly irrelevant \cite{balents}. The static Volovik effect
  is marginally relevant \cite{static,tough} and cause finite density of states at zero energy.
  Surprisingly, the dynamic Volovik effect is highly irrelevant. Note that its coefficient $\Phi^{-2}_{0}$
  diverges as we approach the QCP from the underdoped side.

    In the superconductor phase, the vortex $ \Phi $ is massive, therefore
    can be integrated out, it leads to the old Maxwell term 
    $  \frac{1}{4 m_{\Phi}} ( f_{\mu \nu})^{2} $.
    Integrating out $ a_{\mu} $ generates a mass term  
   $ \frac{1}{2} m_{\Phi}( A^{eff}_{\mu})^{2}_{t} $. From the mass term,
   we can identify the automatically conserved electric current 
   $ I_{\mu}= m_{\Phi} ( J_{\mu} )_{t} $. Note that although the quasi-particle
    electric current is not conserved itself, the total electric current
    $ I_{\mu}= m_{\Phi} ( \delta_{\mu \nu}-\frac{q_{\mu} q_{\nu} }{q^{2}} )
         J_{\nu}(q) $ is conserved. The dynamic Doppler shift effect due to the virtually fluctuating
   vortex-anti-vortex pairs is $ \sim m_{\Phi} ( J_{\mu} )^{2}_{t} $. It is a four fermion interaction
   term, therefore irrelevant. Note that its coefficient vanishes as we approach the QCP from
   the superconducting side.

     We reach the conclusion that the dynamic Doppler shift term is irrelevant on both sides of the QCP.
   We conjecture that it is likely also irrelevant at the QCP. Counter-intuitively,
   it contains two more derivatives in the vortex condensed underdoped regime than in the vortex depleted
   superconducting phase, therefore even more irrelevant in the underdoped regime than in the superconducting
   phase. The previous scaling argument on the irrelevance
    of this term at the QCP presented in Ref.\cite{balents} is questionable, because the authors scaled
   the phase $\phi$ like a free field. In fact, $\phi$ should be confined to be $ 0 < \phi < 2 \pi $, so it 
   is far from being a free field.

    Putting $ \phi_{A}= \phi_{B} = \phi/2 $ in Eqn.\ref{gen1},
   we get the electron annihilation operator:
\begin{equation}
   C_{\alpha}= \sum_{i=1,2} [ e^{i \phi/2 } e^{i \vec{K}_{i} \cdot \vec{x} }
    \psi_{i 1 \alpha} -\epsilon_{\alpha \beta}
    e^{i \phi/2 } e^{-i \vec{K}_{i} \cdot \vec{x} }
    \psi^{\dagger}_{i 2 \beta} ]
\label{sep}
\end{equation}

   The above Eqn. demonstrates clearly that electron is separated
  into charge $ e $ boson $ b= e^{i \phi/2} $ and spin $ 1/2 $ spinon $ \psi $.
  Note that the crucial difference from Eqn.\ref{tough} is that there is {\em no}
  internal gauge field !

   The electron Green function is given by:
\begin{eqnarray}
   & & G(\vec{x}, t) = < b(\vec{x},t) b^{\dagger}(0,0) >
  [  e^{i \vec{K}_{i} \cdot \vec{x}}
    < \psi_{i 1 \alpha}(\vec{x},t)
    \psi^{\dagger}_{i 1 \alpha}(0,0)>   \nonumber  \\
   & + & e^{-i \vec{K}_{i} \cdot \vec{x}}
   < \psi^{\dagger}_{i 2 \alpha}(\vec{x},t) \psi_{i 2 \alpha}(0,0)> ]
\end{eqnarray}

    In the underdoped regime, boson $ b= e^{i \phi/2} $ is massive, therefore
  short-ranged correlated. Being the product of fermion and boson
  Green functions, the electron correlation function is also
  short-ranged. The underdoped regime is a Mott insulator as suggested
  in Refs.\cite{balents,z2,guess}.

  At the critical point, in real space,
  $ b $ decays with a power law $ 1+ \eta_{XY} $,
  the electron Green function decays with a power $ 2+1+\eta_{XY} $.

  In the superconductivity phase, $ <b> \neq 0 $, 
  the electron Green function decays with a power $ 2 $.
  The anomalous Green functions $ F(\vec{x},t)
   = <C_{\uparrow}(\vec{x},t) C_{\downarrow}(0,0)> $ also starts to form:
\begin{eqnarray}
   F(\vec{x}, t) & = & <b>^{2} [ e^{i \vec{K}_{i} \cdot \vec{x}}
    < \psi_{i 1 \uparrow}(\vec{x},t) \psi{\dagger}_{i 2 \uparrow}(0,0)>
      \nonumber  \\
   & - & e^{-i \vec{K}_{i} \cdot \vec{x}} 
   < \psi^{\dagger}_{i 2 \downarrow}(\vec{x},t) \psi_{i 1 \downarrow}(0,0)> ]
\label{super2}
\end{eqnarray}
    It also decays with power 2. Note that the matrix elements between different
   $ p-h $ indices are non-vanishing.

\end{multicols}

\begin{references}
\bibitem{balents} L. Balents, M. P. A. Fisher and C. Nayak,
  Int. J. Mod. Phys. B12, 1033(1998),
   Phys. Rev. B60, 1654 (1999), {\sl ibid} 61, 6307 (2000).
   For a review, see M. P. A. Fisher, cond-mat/9806164.

\bibitem{own} S. Sachdev, Phys. Rev. B44, 380 (1992);
              P. A. Lee and N. Nagaosa, {\sl ibid.} 46, 5621 (1992).

\bibitem{lee} S. H. Simon and P. A. Lee. Phys. Rev. Lett. 78, 1548 (1997).

\bibitem{deri} Jinwu Ye and S. Sachdev, Phys. Rev. B44, 10173 (1991). 
     Starting from the Mott insulating phase, this paper discussed carefully
     the internal uniform and staggered gauge fields. See Eqn.2.12 of this paper.

\bibitem{exchange} Note that under this transformation, the Time (T) reversal
   symmetry of $ c $ becomes Particle and Hole (PH) symmetry of $ d $ and
   vice versa.  In the following, we use the language in the
   $ d $ representation. 

\bibitem{static} Jinwu Ye, Phys. Rev. Lett. 86, 316 (2001).

\bibitem{mac} Y. Wang and A. H. MacDonald, Phys. Rev. B 52, R3876 (1995).

\bibitem{volovik} G. E. Volovik, Sov. Phys. JETP 58, 469 (1993)

\bibitem{hirs} C. Kubert and P. J. Hirschfeld, Phys. Rev. Lett. 80, 
 4963 (1998).

\bibitem{and} P. W. Anderson, cond-mat/9812063.


\bibitem{sing} M. Franz and Z. Tesanovic, Phys. Rev. Lett. 84, 554 (2000). This paper divided static
   vortex lattice into two sublattices.
   For static vortices, the method is equivalent to Anderson's gauge. However,
   as explained in the text, it can not be applied to moving vortices discussed in this paper and in \cite{tough}.
   It can not be used either for a single vortex discussed in \cite{static} and \cite{kov1}. Anderson's original
    gauge transformation can be employed to study all these cases.

\bibitem{bert1} Luca Marinelli, B. I. Halperin  and S. H. Simon, 
   Phys. Rev. B 62, 3488 (2000).

\bibitem{india} A. Vishwanath, cond-mat/0104213.

\bibitem{kov1} A. S. Melnikov, Phys. Rev. Lett. 86, 4198 (2001).

\bibitem{kov2} A. S. Melnikov, J. Phys. Cond. Matt. 82, 4703 (1999).

\bibitem{ong} K. Krishana {\sl et al}, Science 277, 83 (1997),
  Phys. Rev. Lett. 82, 5108 (1999); H. Aubin {\sl et al} Science 280, 9a (1998),
  Phys. Rev. Lett. 624, 5108 (1999).

\bibitem{dual} C. Dasgupta and B. I. Halperin, Phys. Rev. Lett. 47,
  1556 (1981); X. G. Wen and A. Zee, Int. J. Mod. Phys. B 4,437 (1990).

\bibitem{berry} M. P. A. Fisher and D. H. Lee, Phys. Rev. B39, 2756 (1989).

\bibitem{z2} T. Senthil and M. P. A. Fisher, cond-mat/9910224,
    9912380,0006481,0006500,0008082.

\bibitem{guess} C. Lannert, M. P. A. Fisher, T. Senthil, cond-mat/0007002.

\bibitem{u1} L. B. Ioffe and A. L. Larkin, Phys. Rev. B39, 8988 (1989);
    N. Nagaosa and P. A. Lee, Phys. Rev. Lett. 64, 2450 (1990); Phy. Rev. B45, 966 (1992).
   X. G. Wen and P.A. Lee, Phys. Rev. Letts. 80, 2193 (1998);
   P. A. Lee, {\sl et. al.}, Phy. Rev. B 57, 6003 (1998).
   W. Rantner and X. G. Wen, cond-mat/0010378.



\bibitem{break} We can also directly add Eqn.\ref{linear} and
  Eqn.\ref{tilde} without
  performing this singular gauge transformation and then take continuum limit.
  But the resulting continuum action breaks spin $ SU(2) $ symmetry.

\bibitem{magnetic} In the impurity scattering case, the inter-node scatterings maybe
   important because the quasi-particle momenta need not be conserved
   in this case.
   see A. C. Durst and P. A. Lee, Phys. Rev. B 62, 1270 (2000).
   We do not consider the possible magnetic ordering transition connecting
   different nodes in this paper.

\bibitem{hole} If the quasi-particle is hole-like, then $ a_{\mu} 
   \rightarrow -a_{\mu} $, but $ v_{\mu} $ which is gauge invariant
   stays the same. We reach the same conclusion.



\bibitem{wrong} The curvature term in Ref.\cite{sing} is wrong due to
      the ignorance of the gauge invariance.
     

\bibitem{zhang} S. C. Zhang, T. H. Hansen and S. Kivelson, Phys. Rev. Lett.
     62, 82 (1989).

\bibitem{hlr} B. I. Halperin, P. A. Lee and N. Read, Phys. Rev. B47,
    7312 (1993).

\bibitem{typeone} A. K. Geim, S. J. Bending and I. V. Grigorieva, Phys. Rev. Lett.
     69, 2252 (1992).

\bibitem{chen} W. Chen, M. P. A. Fisher and Y. S. Wu, Phys. Rev. B48,
   13749 (1993).

\bibitem{wen} X. G. Wen and Y. S. Wu, Phys. Rev. Lett. {\bf 70}, 1501 (1993).

\bibitem{boson} G. W. Semenoff,  P. Sodano and Y. S. Wu, Phys. Rev. Lett.
  {\bf 62}, 715 (1989); W. Chen, Phys. Lett. B251, 415 (1990).

\bibitem{subir} Jinwu Ye and S. Sachdev, Phys. Rev. Lett, 80, 5409 (1998);
                Jinwu Ye, Phys. Rev. B60, 8290 (1999).

\bibitem{recent} M. Campostrini {\sl et al}, cond-mat/0010360.

\bibitem{qed} T. Appelquist {\sl et al}, Phys. Rev. Lett, 60, 2575 (1988);
          D. Nash, {\sl ibid}, 62, 3024 (1989).

\bibitem{brad} B. Marston, Phys. Rev. Lett. {\bf 64}, 1166 (1990).

\bibitem{kun} Yong Baek Kim and A. J. Millis, Physica, Vol. A4, 171 (1999);
 I. Ussishkin and A. Stern, Phys. Rev. B56, 4013 (1997);
 S. Sakhi, {\sl ibid.} 56, 4098 (1997);
 Kun Yang, {\sl ibid.} 58, R4246 (1998);
 Y. Oreg and B. I. Halperin, {\sl ibid.} 60, 5679 (1999).

\bibitem{double} In double layer Quantum Hall systems, due to the T symmetry breaking
  by the external magnetic field, there is also a conventional CS term term ( the electron
  in a given layer sees its own flux ). The $ K $ matrix \cite{kun} contains a zero
  eigenvalue, it indicates a Goldstone mode which has direct experimental consequences.
  See I. B. Spielman, {\sl et al}, Phys. Rev. Lett, 84, 5808 (2000); cond-mat/0012094.
  In Eqn.\ref{mutual}, the $ K $ matrix only has off-diagonal elements, therefore invertible.

\bibitem{kirtley} J.R. Kirtley {\sl et al}, cond-mat/0103474.

\bibitem{emery} V. J. Emery and S. A. Kivelson, Nature 374, 434 (1995);
                G. Williams,  Phys. Rev. Lett. 82, 1201 (1999).

\bibitem{tough} Jinwu Ye,  Phys. Rev. Lett. 87, 227003 (2001).

\bibitem{feed} B. L. Altshuler and L. B. Ioffe,
   Phys. Rev. Lett, 69, 2979 (1992);
   D. V. Khveshchenko and S. V. Meshkov, Phys. Rev. B 47, 12051 (1993).
   G. Gavazzi, J. M. Wheatley  and A. J. Schofield, Phys. Rev. B 47, 15170
   (1993).

\bibitem{andy} M. Franz and A. Millis, Phys. Rev. B58, 14572 (1998);
 H. J. Kwon and A. T. Dorsey, {\sl ibid} 59, 6438 (1999). For a review, see S.Sharapov, Physics Reports 349, 1 (2001)

\bibitem{kiv} E. Fradkin and S. Kivelson, Nucl. Phys. B474, 543 (1996).

\bibitem{not}  S. Kivelson {\sl et al}, Phys. Rev. B55, 15552 (1997).

\bibitem{coul} M. P. A. Fisher and G. Grinstein,
  Phys. Rev. Lett, 60, 208 (1988); Jinwu Ye, Phys. Rev. B58, 9450 (1998).

\bibitem{sit} M. C. Cha {\sl et al}, Phys. Rev. B44, 6883 (1991);
    M. Wallin {\sl et al}, {\sl ibid} 49, 12115 (1994).
\end{references}
\end{document}